\documentclass[manuscript]{aastex}
\usepackage{color}
\usepackage{amsmath,graphics,graphicx}
\usepackage{rotating}

\newcommand{\lapprox} {\, \lower3pt\hbox{$\sim$}\llap{\raise2pt\hbox{$<$}}\,}
\newcommand{\gapprox} {\, \lower3pt\hbox{$\sim$}\llap{\raise2pt\hbox{$>$}}\,}

\shorttitle{Hard X-ray solar flare loops}
\shortauthors{Jeffrey et al.}

\begin{document}

\title{ON THE VARIATION OF SOLAR FLARE CORONAL X-RAY SOURCE SIZES WITH ENERGY}

\author{Natasha L. S. Jeffrey\altaffilmark{1}, Eduard P. Kontar\altaffilmark{1}, Nicolas H. Bian\altaffilmark{1}, \& A. Gordon Emslie\altaffilmark{2} }

\altaffiltext{1}{School of Physics \& Astronomy, University of Glasgow, G12 8QQ, Glasgow, Scotland, United Kingdom \\
Email: n.jeffrey@physics.gla.ac.uk}

\altaffiltext{2}{Department of Physics \& Astronomy, Western Kentucky University, Bowling Green, KY 42101}

\begin{abstract}
Observations with {\em RHESSI} have enabled the detailed study of the structure of dense hard X-ray coronal sources in solar flares. The variation of source extent with electron energy has been discussed in the context of streaming of non-thermal particles in a one-dimensional cold-target model, and the results used to constrain both the physical extent of, and density within, the electron acceleration region. Here we extend this investigation to a more physically realistic model of electron transport that takes into account  the finite temperature of the ambient plasma, the initial pitch-angle distribution of the accelerated electrons, and the effects of collisional pitch-angle scattering.  The finite temperature results in the thermal diffusion of electrons, that leads to the observationally-inferred value of the acceleration region volume being an overestimate of its true value. The different directions of the electron trajectories, a consequence of both the non-zero injection pitch-angle and scattering within the target, cause the projected propagation distance parallel to the guiding magnetic field to be reduced, so that a one-dimensional interpretation can overestimate the actual density by a factor of up to $\sim 6$.  The implications of these results for the determination of acceleration region properties (specific acceleration rate, filling factor, etc.) are discussed.
\end{abstract}


\section{Introduction}\label{intro}

During a solar flare, the surrounding plasma is heated to tens of mega-Kelvin and electrons are accelerated to deka-keV energies and beyond. In a simple model, electrons travel through a tenuous corona and deposit energy into a dense chromospheric ``thick target'' via Coulomb collisions, with only a small fraction ($\sim 10^{-5}$) of the energy emitted as bremsstrahlung hard X-rays, predominantly at the dense chromospheric footpoints.  Hard X-rays emitted from the corona are usually interpreted as predominantly thermal bremsstrahlung from a hot coronal plasma or as a combination of thermal and thin-target emission.

Over the last decade, the Ramaty High Energy Solar Spectroscopic Imager \citep[{\em RHESSI};][]{2002SoPh..210....3L} has provided unprecedented imaging spectroscopy observations of both chromospheric and coronal X-ray sources
\citep[for recent reviews of this topic see][]{2011SSRv..159..107H,2011SSRv..159..301K}. The design of the {\em RHESSI} instrument is such that spatial information is fundamentally encoded as two-dimensional Fourier transforms, or {\it visibilities}.  The subsequent development of sophisticated and reliable visibility-based image reconstruction algorithms, such as visibility forward-fitting \citep{2002SoPh..210...61H,2007SoPh..240..241S} and \textit{uv\_smooth} \citep{2009ApJ...703.2004M}, coupled with the use of {\it electron visibilities} \citep[spectral inversions of the count visibility data provided by {\em RHESSI};][]{2007ApJ...665..846P} have allowed the quantitative analysis of solar hard X-ray sources in both photon and electron space.

{\em RHESSI} observations have revealed the morphological details of flares with high plasma density \citep[e.g.,][]{1980ApJ...238L..43M,1981A&A....97..210C,1994ApJ...424..444F}, in which the bulk of the hard X-rays come from the corona, with only very weak or no strong footpoint emission from the chromosphere \citep[e.g.,][]{2004ApJ...603L.117V,2004ApJ...612..546S,2007ApJ...666.1256B,2008ApJ...673..576X,2013ApJ...769L..11L}.
The behavior of the source extent with energy is not consistent with a thermal source characterized by a temperature distribution with a peak at the loop-apex, since for such a source, the source size should decrease with energy.  Rather, the source extent {\it grows} with energy \citep{2008ApJ...673..576X}, indicative of a nonthermal model in which the propagation distance increases with energy. Apparently, the density within the coronal region in such sources is high enough to stop electrons prior to reaching the chromosphere; the source is a coronal ``thick target''.

Studying these events is particularly valuable since: (1) the coronal X-ray component and hence acceleration region can be studied without contamination from an intense chromospheric source; and (2) such sources exhibit trends in source extent with energy \citep{2008ApJ...673..576X, 2011ApJ...730L..22K, 2012ApJ...755...32G,2013ApJ...766...28G} and time \citep{2013ApJ...766...75J}, which can be used to study particle acceleration and transport processes \citep[e.g.,][]{2012SoPh..277..299G,2013SoPh..284..489G}.  Further, unlike footpoint-dominated solar flares \citep[e.g.,][]{1982SoPh...78..107A,1982SoPh...81..137D,1995PASJ...47..355T,1996AdSpR..17...67S,1999ApJ...527..945P,2003ApJ...595L.107E,2004A&A...415..377M,2007A&A...461..315T,2011A&A...533L...2B,2011ApJ...731L..19F,2013ApJ...777...33C},
the HXR spectra of such ``coronal thick-target sources'' tends to be softer than, and the sources higher than, chromospheric sources, which generally reduces the albedo contribution to X-ray images \citep{2010A&A...513L...2K}, making the interpretation of the spectro-spatial structure of such sources more straightforward.

Observations of compact coronal nonthermal hard X-ray sources typically show that the extent of the source parallel to the guiding magnetic field increases approximately quadratically with photon energy. Since the collisional stopping distance of an electron in a plasma also increases quadratically with particle energy, \cite{2008ApJ...673..576X} explained this behavior in terms of an extended acceleration region, from which accelerated electrons emerge and subsequently undergo collisional transport in a background medium of uniform density.  As shown by \citet{2008AIPC.1039....3E}, application of such a model allows parameters such as the number density $n$ of the region and the specific electron acceleration rate $\eta$ (electrons~s$^{-1}$ per ambient electron) to be estimated.

However, the simple one-dimensional cold-target approximation used by these authors is not completely realistic, for two main reasons. First, it assumes that the injected electron trajectories are completely aligned with the guiding magnetic field, and it does not take into account pitch-angle scattering (collisional or otherwise) of the accelerated electrons in the target. Second, it neglects effects associated with the finite temperature of the ambient medium; electrons with energies comparable to the thermal energy of the plasma $\sim k_BT$ are just as likely to gain as lose energy during a collision, unlike the monotonic energy loss experienced by suprathermal electrons interacting with a cold plasma \citep[e.g.,][]{1978ApJ...224..241E}.  Even for electrons that do lose energy, they do so at a rate that is not the same as in a cold-target, so that a quadratic behavior of source extent with energy is not necessarily expected.

\cite{2003ApJ...595L.119E} and \cite{2005A&A...438.1107G} investigated analytically the effects of a finite target temperature, and both found that the associated velocity diffusion cannot be neglected when interpreting the results of flare hard X-ray spectra. \cite{2003ApJ...595L.119E} found that, because of the reduced energy losses suffered by accelerated electrons in  warm-target, the inferred energy content of the injected electron distribution was significantly reduced. Indeed, he showed that allowance for this effect obviated the need to introduce a low-energy cutoff in the electron distribution. \cite{2005A&A...438.1107G} found that changes occurring close to the thermal energy of the plasma meant that many flare X-ray spectra may not be well fitted by a simple isothermal-plus-power-law model.

The main motivation of our study is to incorporate the effects of both finite target temperature and non-zero pitch-angle (due to both the finite width of the injected pitch-angle distribution and scattering within the target) in models of the variation of source size with electron energy.  We investigate how the inclusion of each of these processes changes the behavior of the variation of source extent with electron energy, and in turn how this affects the estimation of parameters such as number density $n$ and acceleration region length $L_{0}$. We also briefly discuss how the values of inferred parameters such as the filling factor $f$ and specific electron acceleration rate $\eta$ are changed by the inclusion of such processes.

\section{Electron collisional transport in a cold plasma}\label{cold_theory}

We first briefly review electron transport within a uniform cold-target (i.e., electron energy $E>>k_BT$, where $k_B$ is Boltzmann's constant and $T$ is the target temperature), ignoring the effects of collisional pitch-angle scattering. The variation of energy $E$ (erg) with position $z$ (cm)\footnote{For easy comparison with solar observation, we will present results in arcseconds where $1''=7.25\times10^7$~cm at the Sun.} in such a model is given by \citep[cf.][]{1978ApJ...224..241E}

\begin{equation}\label{eq:et}
E(E_{0},z)=\sqrt{E_{0}^{2}-2KN(z)}
=\sqrt{E_{0}^{2}-2Kn \, |z-z_{0}|} \,\,\, ,
\end{equation}
where $z_{0}$ is the (single) point of injection, $K=2\pi e^{4} \ln\Lambda$ (where $e$ (esu) is the electron charge and $\ln\Lambda$ the Coulomb logarithm), and $N$ and $n$ are the column density (cm$^{-2}$) along the trajectory and ambient number density (cm$^{-3}$), respectively.

This expression allows us to find the stopping position $L_{S}$ of an electron of initial energy $E_{0}$ within a plasma of density $n$ \citep[cf.][]{2002SoPh..210..373B}, viz.

\begin{equation}\label{eq: sd2}
L_{S}=z_{0}+\frac{E_{0}^{2}}{2Kn} \,\,\, .
\end{equation}
Using Equation~(\ref{eq:et}) and the one-dimensional continuity equation, we can also obtain the form of the electron spectrum as a function of position in the target:

\begin{equation}\label{eq: fex}
F(E,z)=F_{0}(E_0) \, \frac{dE_{0}}{dE} = F_{0}(E_0) \, \frac{E}{E_0}
= \frac{E}{\sqrt{E^{2}+2Kn|z-z_{0}|}} \, F_{0}(E_0[E,z]) \,\,\, .
\end{equation}
Setting $z_{0}=0$ and assuming a power-law injection spectrum $F_{0}(E_0) \propto E_{0}^{-\delta}$, we can derive an expression for the source extent $\sigma$ as the square root of the variance

\begin{equation}\label{eq: varfex1}
\sigma^2(E)=\frac{\int_{0}^{\infty} z^2 \, (E^{2}+2Knz)^{-(\delta+1)/2} \, dz}
{\int_{0}^{\infty} (E^{2}+2Knz)^{-(\delta+1)/2} \, dz} \,\,\, ,
\end{equation}
where the symmetry about $z=0$ has been used.  Evaluating the integrals gives

\begin{equation}\label{eq:stdfex}
\sigma(E)
=\frac{1}{2Kn} \, \sqrt{\frac{8}{(\delta-3)(\delta-5)}} \, E^{2} \,\,\, .
\end{equation}
The spatial extent at a given energy $E$ depends on the spectral index $\delta$; for $\delta=7$, we obtain the form of the stopping distance $\sigma=L_{s}$ given by Equation~(\ref{eq: sd2}). It should be noted that Equation (\ref{eq: varfex1}), and hence the spatial extent defined by Equation~(\ref{eq:stdfex}), is applicable only for $\delta >5$; for $\delta \leq 5$, the integral on the numerator diverges at the upper limit. This is related to the fact that the collisional stopping length is an increasing function of energy $\propto E^2$, so that large energies give the largest contribution to the integral for $\delta \leq 5$.  This issue is formally avoided by imposing an upper energy cut-off $E_{max}$ to $F_0(E_0)$, so that the upper limit in the integral (\ref{eq: varfex1}) is finite, given by $E_{max}^2/2Kn$.

\begin{figure*}
\centering
\includegraphics[width=16cm]{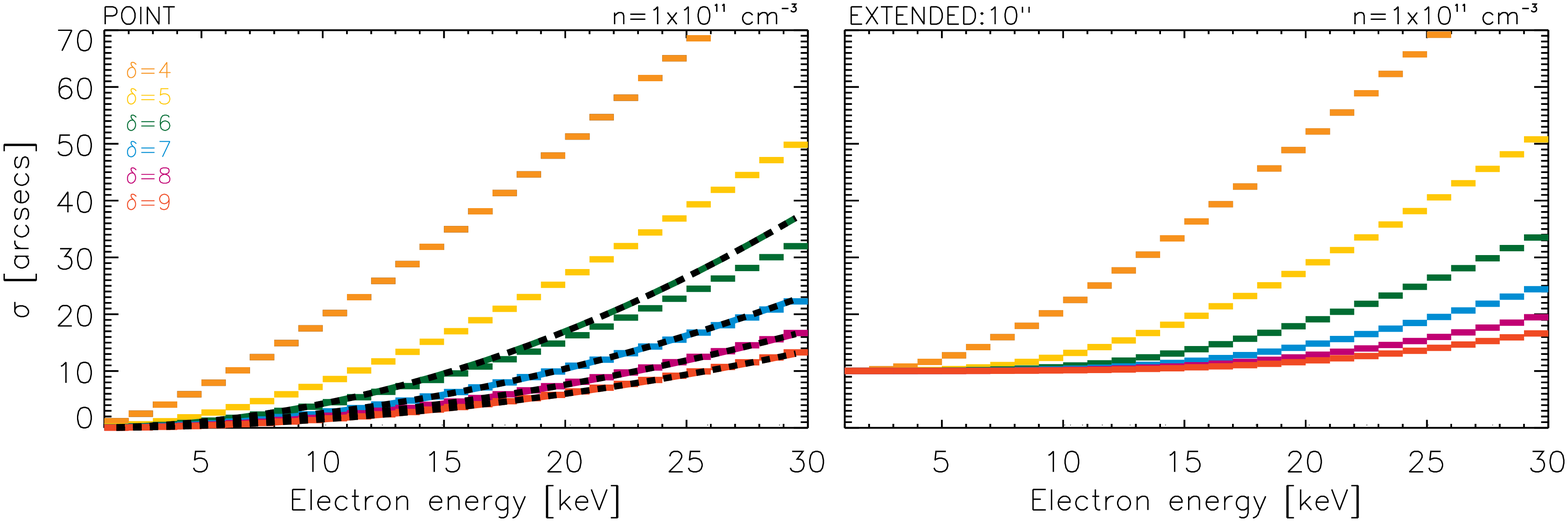}
\includegraphics[width=16cm]{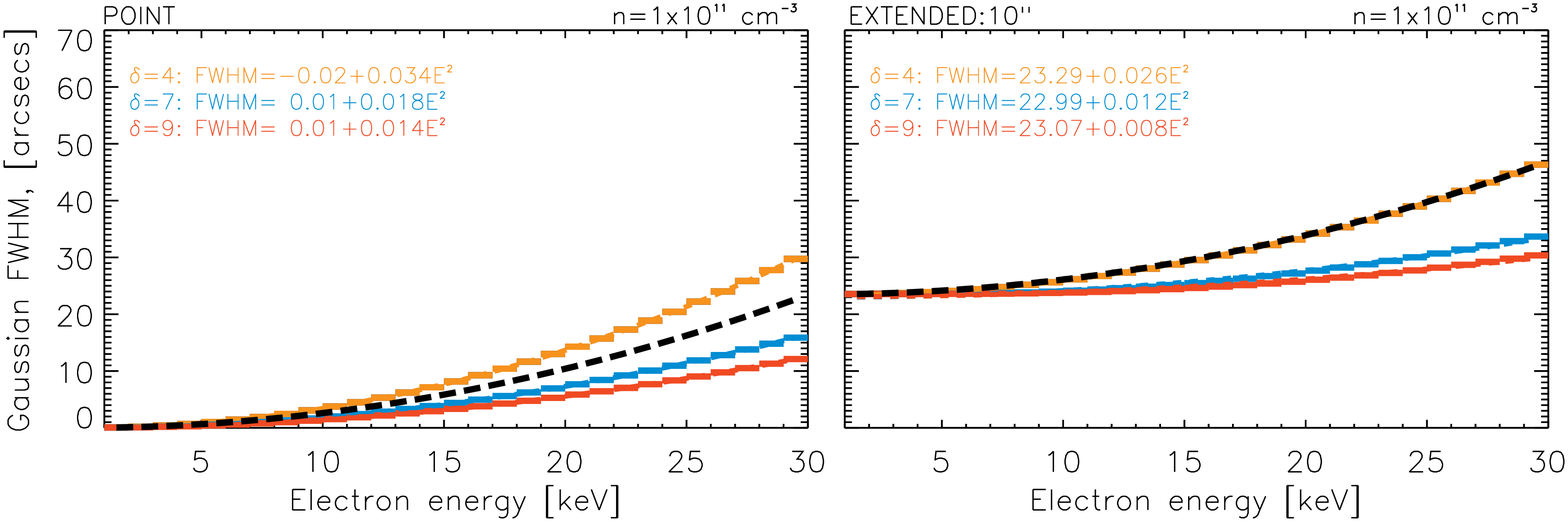}
\caption{{\it Top panels}: the standard deviation $\sigma$ calculated for a point source (left) and a source of Gaussian standard deviation $d=10\arcsec$ (right), using the moment-based Equation~(\ref{eq: varfex1}) and the distribution of electron flux with energy and position given by Equation~(\ref{eq:fexgauss}), for a target density $n=1\times10^{11}$~cm$^{-3}$. For the point source, the curves calculated using Equation~(\ref{eq:stdfex}) for $\delta=6-9$ and a maximum injected energy of $30$~keV are over-plotted as dashed lines of the same color. {\it Bottom panels}: Gaussian FWHM calculated by fitting Gaussian curves to $F(E,z)$ for a point source (left) and a $10''$ source (right). { Equation~(\ref{L}) is fitted to each curve:  the corresponding values of $L_0$ and $\alpha$ are shown on each panel. The curve FWHM $=2 \, \sqrt{2 \ln 2} \, d + E^2/2Kn$ \citep[black dashed curve; used by previous authors - e.g., ][]{2012A&A...543A..53G} is also superimposed.}}
\label{fex_good}
\end{figure*}

If the initial electron distribution is injected over a finite region, with the injected flux profile having the form of a Gaussian distribution with standard deviation $d$, then the equation  for $F(E,z)$ becomes \citep[see, e.g.,][]{2014ApJ...780..176K}

\begin{equation}F(E,z)\label{eq:fexgauss}
\sim \frac{1}{d\sqrt{2\pi}} \, \int_{-\infty}^{\infty} E \, \left(E^{2}+2Kn \, \vert z-z^{'}\vert \right)^{-(\delta+1)/2}
\exp{\left(-\frac{{z^{'}}^2}{2d^{2}}\right)} \, dz^{'} \,\,\, .
\end{equation}
For this case, the evaluation of $F(E,z)$ and the corresponding standard deviation $\sigma(E)$ cannot be evaluated analytically. Figure~\ref{fex_good} (top) shows the numerical results for $\sigma(E)$ for $\delta=4-9$ using initial source sizes of $d=0\arcsec\;\mbox{and}\;10\arcsec$ and a number density $n_{0}=1\times10^{11}$~cm$^{-3}$. For the $d=0\arcsec$ case, and for cases with $\delta > 5$ (cf. Equation~(\ref{eq:stdfex})), the $\sigma(E)$ results calculated from the point-injection case~(Equation~(\ref{eq:stdfex})) are over-plotted for comparison.

The form of the spatially resolved spectrum $F(z)$ at a given energy $E$ at distances further away from the peak, where $F(E,z) \lapprox 0.15\max[F(E)]$ (cf. Equation~(\ref{eq:fexgauss})) is not well determined by the {\em RHESSI} observations.
Thus there is considerable value in calculating $\sigma$ not through a moment-based approach, but rather through a shape-based analysis that focuses\footnote{Keeping in mind the fundamental nature of {\em RHESSI} data as spatial Fourier transforms (visibilities), we note that the source sizes in practice are determined by fitting a Gaussian-like shape to the observed visibilities. Due to this indirect imaging approach and the finite dynamic range of the instrument, the brightest part of the image is the most reliable.} on the high-intensity ``core'' of the spatial distribution of flux at a given electron energy $E$.  Therefore, we also fitted Gaussian curves to $F(E,z)$ in order to determine the Gaussian standard deviation $\sigma_{G}$ at each energy. This allows us to calculate the Gaussian Full Width at Half Maximum FWHM$=2 \, \sqrt{2 \ln 2} \, \sigma_{G}$. The curves for $\delta= $4, 7, and 9 are plotted in the bottom panels of Figure~\ref{fex_good}. In general, and as expected, the $\sigma_G(E)$ values deduced from the shape of the core of the $F(E,z)$ profile are smaller than the $\sigma(E)$ values deduced from the moment-based analysis. Each curve in Figure~\ref{fex_good} (bottom) was fitted with an equation of the form

\begin{equation}
\mbox{FWHM}(E)=L_{0}+\alpha E^{2}
\label{L}
\end{equation}
and the values of $L_{0}$ and $\alpha$ are shown on each plot. In the bottom panels of Figure~\ref{fex_good},  we have also over-plotted FWHM$=2 \, \sqrt{2 \ln 2} \, d + E^2/2Kn$ for comparison, since this simple approximation has been used \citep[e.g.,][]{2012A&A...543A..53G} to infer information from observations; it is given by the black dashed curve.

From Figure~\ref{fex_good}, we note two main points:

\begin{enumerate}
\item{For a given energy $E$, $\sigma(E)$ decreases with increasing spectral index $\delta$.  This is because as $\delta$ increases there are a lower proportion of higher energy electrons in the overall electron distribution.  The lower energy electrons that are representative of steeper spectra travel a smaller distance through the plasma.}
\item{For a given spectral index $\delta$, the value of the quadratic coefficient $\alpha$ decreases somewhat with source size $d$. This is because of the increased contribution of the acceleration region to the overall source extent; the ``propagation'' region is to a large extent contained within the acceleration region itself.}
\end{enumerate}
Observationally, $L_{0}$ is used to infer the size of the acceleration region, while $\alpha\propto 1/n$ allows us to infer the number density of the propagation region (assumed to be the same as the density of the acceleration region). Using the simplest one-dimensional cold plasma approximation ($\alpha=1/2Kn$), $n$ can be inferred easily. However, from Figure~\ref{fex_good}, we can see that in general $\alpha=\beta/2Kn$, where the value of the dimensionless number $\beta$ {and hence the number density $n$, depends upon the properties of both the acceleration region and the electron distribution.}

Further, Equations~(\ref{eq:stdfex}) and~(\ref{eq:fexgauss}) do not account for three important processes we would expect to occur within a real flaring coronal plasma: (1) a finite range of pitch-angles in the injected pitch-angle distribution, (2) any form of pitch-angle scattering (collisional or non-collisional) within the target, and (3) the finite temperature of the plasma through which the electrons travel.  All of these physically important effects impact the form of $E(E_0,z)$, the variation of electron energy with position in the source, and incorporating them will thus change the resulting forms of $\sigma(E)$ and FWHM$(E)$, in a manner which we now proceed to investigate.

\section{Electron transport in a {\bf finite temperature Maxwellian} plasma with collisional pitch-angle scattering}

As shown in the Appendix, the equation describing the spatial distribution of the electron flux spectrum $F(E,\mu,z)$ (electrons~cm$^{-2}$~s$^{-1}$~keV$^{-1}$) as a function of field-aligned coordinate $z$ (cm), energy $E$ (keV) and pitch-angle cosine $\mu$ is

\begin{eqnarray}\label{eq: fp_e_zefftext}
\mu \, \frac{\partial F}{\partial z} &= &
\Gamma_{eff} \left \{ \frac{\partial}{\partial E} \left [ G \left (\sqrt{\frac{E}{k_B T}} \, \right ) \, \frac{\partial F}{\partial E} +\frac{1}{E} \, \left ( \frac{E}{k_B T} - 1 \right ) \, G \left (\sqrt{\frac{E}{k_B T}} \, \right ) \, F \right ] \right .
+ \cr
&+ & \left . \frac{1}{8E^2} \, \frac{\partial}{\partial \mu} \left [ (1-\mu^{2}) \, \left ( {\rm erf} \left ( \sqrt{\frac{E}{k_B T}} \, \right ) -G\left (\sqrt{\frac{E}{k_B T}} \, \right ) \right ) \frac{\partial F}{\partial \mu} \right ]\right \} + S(E,\mu,z) \,\,\, ,
\end{eqnarray}
where $T$ is the ambient temperature (K), $k_B$ is Boltzmann's constant, $\Gamma_{eff}$ is a Coulomb coefficient defined in the Appendix and $G(u)$ is the Chandrasekhar function, given by

\begin{equation}\label{eq:gchatext}
G(u)=\frac{{\rm erf}(u)-u \, {\rm erf}^{'}(u)}{2u^{2}} \,\,\, ,
\end{equation}
where ${\rm erf}(u)\equiv (2/\sqrt{\pi})\int\limits_{0}^{u}\exp(-t^2) \, dt$ is the error function. $S(E,\mu,z)$ is a source term which we assume to be of separable form in $E,\,\mu$ and $z$:

\begin{equation}\label{eq:source}
S(E,\mu,z) = F_{0}(E) \; \frac{1}{\sqrt{2\pi d^{2}}} \, \exp{\left(-\frac{z^{2}}{2d^{2}}\right)} \, H(\mu) \,\,\, ,
\end{equation}
where $F_0(E)$ and $H(\mu)$ describe the forms of the initial energy spectrum and pitch-angle distribution, respectively. Equation~(\ref{eq: fp_e_zefftext}) describes the evolution of an injected electron distribution through a non-evolving finite temperature background Maxwellian distribution. Ignoring the source term for now in order to focus on the electron transport, Equation~(\ref{eq: fp_e_zefftext}) can be written in the form

\begin{equation}\label{eq: fp_eABin}
\mu \, \frac{\partial F}{\partial z}
=\frac{\partial}{\partial E}\biggl ( A_E(E) \, F \biggr )
+ \frac{\partial^{2}}{\partial E^{2}} \biggl ( D_{EE}(E) \, F \biggr )
+\frac{\partial}{\partial \mu} \biggl ( A_\mu(E,\mu) \, F \biggr )
+\frac{\partial^2}{\partial \mu^2} \biggl ( D_{\mu\mu}(E,\mu) \, F \biggr )  \,\,\, ,
\end{equation}
where the coefficients
\begin{align}\label{eq:A-D}
A_E(E) &=&  \frac{\Gamma_{eff}}{2E} \left [ {\rm erf} \left (\sqrt{\frac{E}{k_B T}} \, \right ) -2 \sqrt{\frac{E}{k_B T}} \, {\rm erf}^{'} \left (\sqrt{\frac{E}{k_B T}} \, \right ) \right ] \,\,\, && \cr
&\equiv& \frac{\Gamma_{eff}}{2E} \, g_{th} \left (\sqrt{\frac{E}{k_B T}} \right )\,\,\, &;& \cr
D_{EE}(E)\equiv\frac{1}{2} \, B_E^2(E) &=&   \, \Gamma_{eff} \, G \left (\sqrt{\frac{E}{k_B T}} \, \right )  \,\,\, &;&
\cr
A_\mu(E,\mu) &=& \frac{\mu \, \Gamma_{eff}}{4E^2} \left [ {\rm erf} \left ( \sqrt{\frac{E}{k_B T}} \, \right ) -G \left ( \sqrt{\frac{E}{k_B T}} \, \right ) \right ] \,\,\, &;&
\cr
D_{\mu\mu}(E,\mu)\equiv\frac{1}{2} \, B_\mu^2(E,\mu) &=&
 \frac{(1-\mu^{2}) \, \Gamma_{eff}}{8 E^2} \, \left [ {\rm erf}\left (\sqrt{\frac{E}{k_B T}} \, \right ) - G \left (\sqrt{\frac{E}{k_B T}} \, \right ) \right ] \,\,\, &.&
\end{align}

This general form of the Fokker-Planck equation is equivalent to the following stochastic differential equations (SDE) for $E$ and $\mu$ in the It\^o form

\begin{equation}\label{eq:sto_Emu}
{dE}= - \, A_E \, ds + B_E \, \, dW_{E}
 \,\,\, ; \qquad {d\mu}= - \, A_\mu \,ds  + B_\mu \, \, dW_{\mu} \,\,\, ,
\end{equation}
where the independent Wiener processes $W_{\mu}$ and $W_{E}$ are stochastic processes with independent increments. These two equations suggest the numerical stepping algorithm

\begin{figure*}
\centering
\includegraphics[width=17cm]{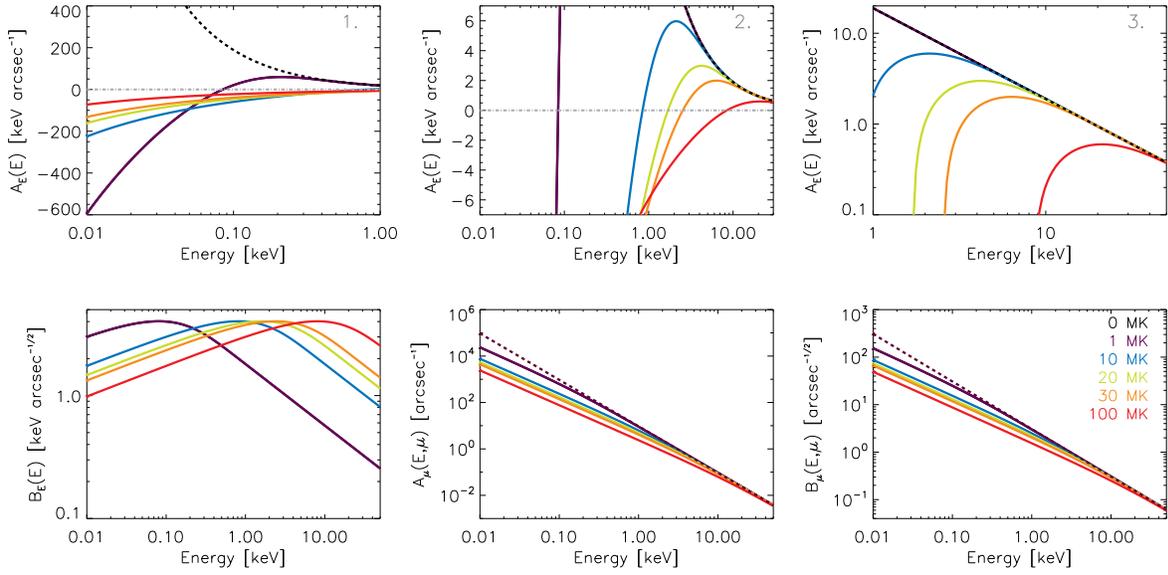}
\caption{Plots of the energy $A_E$, $B_{E}$ and pitch-angle $A_\mu(E,\mu=1)$, $B_{\mu}(E,\mu=0)$ coefficients against electron energy $E$ for different plasma temperatures from $T=0 - 100$~MK. The colors corresponding to each plasma temperature are shown on the bottom right plot. The $A_E$ coefficient is plotted three times (top row) so that all the features in different energy ranges can be seen clearly.}
\label{fig:check_ABCD}
\end{figure*}

\begin{equation}\label{eq:sto_x}
z_{j+1}=z_{j}+\mu_{j} \, \Delta s \,\,\, ;
\end{equation}
\begin{equation}\label{eq:sto_E}
E_{j+1}=E_{j}-\frac{\Gamma_{eff}}{2E_{j}} \, g_{th}(u_j) \, \Delta s
+\sqrt{2 \, \Gamma_{eff} \, G(u_{j}) \, \Delta s} \,\, W_{E} \,\,\, ; \,\,\,
\end{equation}
\begin{equation}\label{eq:sto_mu}
\mu_{j+1}=\mu_{j}-\frac{\Gamma_{eff} \biggl ( {\rm erf}(u_{j})-G(u_{j}) \biggr ) } {4 E_{j}^{2}} \,\, \mu_{j} \, \Delta s
+\sqrt{\frac{ (1-\mu_{j}^{2}) \, \Gamma_{eff} \, \biggl ( {\rm erf}(u_{j})-G(u_{j}) \biggr ) } {4 E_{j}^{2}} \, \Delta s} \, \, W_{\mu} \,\,\, ,
\end{equation}
where $u_j=\sqrt{E_j/k_B T}$ and $W_E$ and $W_{\mu}$ are drawn at random from the Gaussian distribution $N(0,1)$ such that $\langle W_{\mu}\rangle=\langle W_{E}\rangle  =0$, $\langle W^2_{\mu}\rangle=\langle W^2_{E}\rangle= 1$. Equations~(\ref{eq:sto_x}) through~(\ref{eq:sto_mu}) are the form of the SDEs we use\footnote{We took the rms atomic number $\zeta_{eff}=1$,  in $\Gamma_{eff}$ (see Appendix), for simplicity (i.e., we use a pure Hydrogen target), but have provided the equation for a general $\zeta_{eff}$, because it may prove useful in other studies.} in our numerical simulations, which were amended for low energies, see Section~ \ref{low-energy-limit}.

The coefficients $A_E$, $A_\mu$, $B_E (=\sqrt{2D_{EE}})$ and $B_\mu (=\sqrt{2D_{\mu\mu}})$ are plotted against energy $E$ in Figure~\ref{fig:check_ABCD}, for a number of different plasma temperatures $T$ ranging from $T = 0$ (cold plasma) to $T=100$~MK. For ease of presentation, the $A_\mu$ and $B_\mu$ terms are shown as a function of $E$ for a fixed value of $\mu$ ($\mu=1$ for $A_\mu$ and $\mu=0$ for $B_\mu$). Below an energy $E_c\simeq k_BT$ the coefficient $A_E$ becomes negative; i.e., electrons on average {\it gain} energy; the value of $E_c$ for which $A_E=0$ increases linearly with the ambient temperature. In order that these features can be seen clearly, the coefficient $A_E$ is plotted (top row of Figure~\ref{fig:check_ABCD}) over three different energy ranges: two (0.01-1~keV, and 0.01-30~keV) plotted on linear $y$-axes and 1-50~keV plotted on a logarithmic $y$-axis.  Further, the stochastic term $B_{E}$ peaks at $\simeq k_B T$. Therefore, in a warm plasma, electrons with $E \sim k_B T$ are more likely to gain energy, both secularly and through diffusion, rather than to lose it.

To get reliable results from the simulations, an appropriate value of the length step $\Delta s$ (Equations~(\ref{eq:sto_x}) through~(\ref{eq:sto_mu})) must be chosen. This was done by calculating the thermal collision length (mean free path) $\lambda_c(E)$ and ensuring that $\Delta s$ was much smaller than $\lambda_c$ for all $E$ of interest. The mean-free path $\lambda_c$ for a 1~keV electron in a cold-target of density $n=1 \times 10^{11}$~cm$^{-3}$ is approximately $10^6$~cm; the mean-free paths in warm-targets are even greater.  For all our simulations, we use a length step $\Delta s=1\times10^{5}$~cm, much smaller than the mean free path in all cases.

\subsection{The low-energy limit}\label{low-energy-limit}

As the plots in Figure~\ref{fig:check_ABCD} show, $A_E$, $A_\mu$ and $B_\mu$ diverge as $E\rightarrow 0$. Therefore, following \citet{2009JCoPh.228.1391L} and \citet{2010ITPS...38.2394C}, for low energies $E$ we replace the finite difference equation~(\ref{eq:sto_E}) with an analytic expression for the energy evolution. To obtain this expression, the functions ${\rm erf}(u)$ and ${\rm erf'}(u)$ for small $u$ are expanded in a MacLaurin series, so that the coefficients $A_E$ and $B_E$ become

\begin{equation}\label{low_e_A}
A_E = \frac{\Gamma_{eff}}{2E} \left ( {\rm erf}(u)-2u \, {\rm erf}^{'}(u) \right )
\simeq - \, \frac{\Gamma_{eff}}{\sqrt{\pi}E} \, u \,\,\, ,
\end{equation}

\begin{equation}\label{low_e_B}
B_E = \sqrt{2 \, \Gamma_{eff} \, G (u)}
\simeq \sqrt{\frac{4 \, \Gamma_{eff}}{3 \, \sqrt{\pi}} \, u} \,\,\, ,
\end{equation}
and the SDE for energy in the low-energy limit, $E\rightarrow 0$, becomes

\begin{equation}\label{eq:low_de_sde}
\frac{dE}{ds} \simeq \frac{\Gamma_{eff}}{E} \, \sqrt{\frac{E}{\pi k_{B} T}} + \left ( \frac{4 \, \Gamma_{eff}}{3} \, \sqrt{\frac{E}{\pi k_B T}} \, \right )^{1/2} W^{E} \,\,\, .
\end{equation}
For low values of $E$, the second (stochastic) term can be neglected in comparison with the first (secular) term to give

\begin{equation}\label{eq:low_de_sde_A}
\frac{dE}{ds}\simeq \frac{\Gamma_{eff}}{\sqrt{\pi k_{B} T}} \, \frac{1}{\sqrt{E}} \,\,\, ,
\end{equation}
which can be integrated analytically, giving

\begin{equation}\label{eq:low_e_sde_A_2}
E=\left[E_{0}^{3/2}+\frac{3 \, \Gamma_{eff}}{2\sqrt{\pi k_{B} T}} \, (s-s_0)\right]^{2/3} \,\,\, .
\end{equation}
Equation (\ref{eq:low_e_sde_A_2}) was used for energies below

\begin{equation}
E_{low} = \left[\frac{3\Gamma_{eff}}{2\sqrt{\pi k_{B} T}} \, \Delta s  \right]^{2/3}\,\,\, ,
\end{equation}
thus guaranteeing that $E \ge 0$ everywhere. To avoid divergence, the pitch-angle cosine $\mu$ for energies $E\leq E_{low}$ was sampled from a uniform distribution between $-1$ and $1$.

In the cold plasma limit $T\rightarrow0$, the stochastic equation for $E$ becomes

\begin{equation}\label{eq:cold_E}
E_{j+1}=E_{j}-\frac{\Gamma_{eff}}{2 E_{j}} \, \Delta s \,\,\, ,
\end{equation}
which can be solved to give the usual cold-target result

\begin{equation}
E_{j+1}=\sqrt{E^2_{j} - 2 K n \, \Delta s} \,\,\, ,
\end{equation}
where $K=2\Gamma_{eff}/n$. In this limit, the pitch-angle behavior is given by

\begin{equation}\label{eq:cold_mu}
\mu_{j+1}=\mu_{j}-\frac{\Gamma_{eff}}{4 E_{j}^{2}} \, \mu_{j} \, \Delta s + \sqrt{\frac{\Gamma_{eff}}{4 E_{j}^2} \, (1-\mu^{2}) \, \Delta s} \,\, W_{\mu}\,\,\, .
\end{equation}

\section{Simulations}

The aim of our simulations is to determine how collisional pitch-angle scattering and the finite temperature of the target plasma affect the transport of electrons through the plasma compared to the one-dimensional cold-target result, and hence to determine how the observed length of a hard X-ray source varies with electron energy in a more realistic physical scenario.  Our simulations used the stochastic equations for $z$, $E$, and $\mu$ given by Equations~(\ref{eq:sto_x}) through~(\ref{eq:sto_mu}) with initial conditions for each injected electron provided by sampling the source term $S(E,\mu,z)$ -- see Equation (\ref{eq:source}).  Our simulations model the evolution of an injected distribution of electrons, moving either within a cold plasma or a plasma of finite temperature.  They do not account for the evolution of the background plasma; the properties of the background plasma remain constant throughout a simulation.

\subsection{Simulation input, boundary and end conditions}\label{sec:input}

All of our simulations used a common set of certain input parameters.  The electron number density was set to $n = 1 \times 10^{11}$~cm$^{-3}$, a relatively high value for the coronal density, but one which is necessarily high in order for the deka-keV electrons to be stopped in the corona and which is chosen to correspond to recent analyses of thick-target coronal sources \citep[e.g.,][]{2008ApJ...673..576X,2011ApJ...730L..22K,2013ApJ...766...75J}.
For the Coulomb logarithm we used a typical coronal value of $\ln{\Lambda}=20$. The plasma temperature is assumed uniform along the $z$ direction, at a value of either $0$~MK, $10$~MK, $20$~MK or $30$~MK. The initial spatial distribution of injected flux (``acceleration region size'') is assumed to be a Gaussian centered at $z=0$ (e.g., the position of the coronal loop apex) with an input standard deviation of $d=10\arcsec$, corresponding to a FWHM$=2\sqrt{2\ln 2} \, d=23\arcsec.5$.  We took the initial pitch-angle distribution to be either {\it completely beamed} (i.e., half the distribution has $\mu=1$ and the other half $\mu=-1$) or {\it isotropic}. The injected electron energy flux distribution $F_0(E)$ has the form of a power law with spectral index $\delta=4$ or $\delta=7$, up to a maximum energy of $50$ keV, above which the energy-integrated electron flux is negligibly small.

For the runs that use the cold-target energy loss formula, electrons lose energy monotonically.  Hence an electron is removed from the simulation
once its  energy is below $1$~keV and the simulations are terminated when all electrons are removed.
Electrons in the warm-target runs are {\it never} removed for the warm-target simulations, as the electrons of
energy $\simeq k_B T$ can still gain energy through Coulomb collisions with more energetic neighbors,
as the ensemble evolves to a thermal (Maxwellian) distribution.  Thus
for such runs the particle number is conserved and the electron distribution asymptotically approaches
the Maxwellian distribution $F(E)\sim E \exp (-E/k_B T)$.  For this distribution, the flux-averaged energy is

\begin{equation}\label{average-energy}
{\overline E} = \frac {\int^{\infty}_{0} E \, F(E) \, dE}  {\int^{\infty}_{0} F(E) \, dE} = 2 \, k_{B} T \,\,\, .
\end{equation}
The simulation are terminated when the average energy of the distribution is $2 k_B T$ and the pitch-angle distribution
becomes approximately isotropic, conditions that approximate the essential features of a Maxwellian.
Note that $\overline E$ is {\it not} the average kinetic energy of the three-dimensional phase space distribution $f(v,\mu,z)$ (which is $\langle{mv^{2}}/{2}\rangle=\frac{3}{2}k_{B}T$). After each distance step $\Delta s$, the values
of electron distribution function $F(E,\mu,z)$ are saved into an array. These arrays represent the distribution functions
resulting from the continuous injection of electrons with the source function given by Equation (\ref{eq:source}).

\subsection{Gaussian fitting and the determination of the source length FWHM}

The arrays generated by each simulation were energy-binned to give $F(\mu,z)$ in 1~keV energy bins from $1$~keV to $30$~keV. The longitudinal extent of the source could be identified as the standard deviation $\sigma$ of the $F(\mu,z)$ spatial distribution in each energy bin, calculated from the second spatial moment of $F(E,z)$.  However, in part because the injected flux distribution is {\it assumed} to be Gaussian, the forms of $F(\mu,z)$ generally also closely resemble Gaussian forms, excluding relatively low-intensity components at high $|z|$.  Therefore, just as in Section~\ref{cold_theory}, we instead chose to fit a Gaussian distribution to each $F(\mu,z)$ distribution and to thus determine the associated standard deviations $\sigma_{G}$ and corresponding FWHM$=2\sqrt{2 \ln 2} \, \sigma_{G}$ in each energy bin. In this way, we characterize the extent of the source through the {\it shape} of its core spatial form, rather than through a {\it moment} of the entire distribution. Again, as in Section~\ref{cold_theory}, FWHM$(E)=L_{0}+\alpha E^{2}$ (Equation~(\ref{L})) was fitted to the FWHM versus electron energy results, and values of $\alpha$ and $L_0$ found.

For a cold plasma with an initially beamed pitch-angle distribution and no collisional pitch-angle scattering, we would expect $L_{0}=L_{init}=2\sqrt{2 \ln 2} \, d$, the Gaussian FWHM of the input distribution, and a value of $\alpha$ equal to that found numerically from the fits to $\delta=4$ and $\delta=7$ curves in Figure~\ref{fex_good}. However, the presence of a finite plasma temperature $T$, an initially broad pitch-angle distribution, and/or collisional pitch-angle scattering will all change the values of $L_{0}$ and $\alpha$ obtained. The inferred values of the acceleration region density $n$ depend on the value of $\alpha$ ($\alpha \propto 1/n$).  The values of other parameters inferred from $n$ and the acceleration region length $L_0$ -- see Section~\ref{discussion} -- are thus dependent upon both the assumed electron distribution and the properties of the target plasma. We will use our results to find, for instance, how the inappropriate use
of a one-dimensional cold-target assumption changes the inferred number density by a factor larger
than the observational uncertainty, and thus determine if a correction should be applied when X-ray observations are interpreted. 

\begin{figure*}
\centering
\includegraphics[width=75mm]{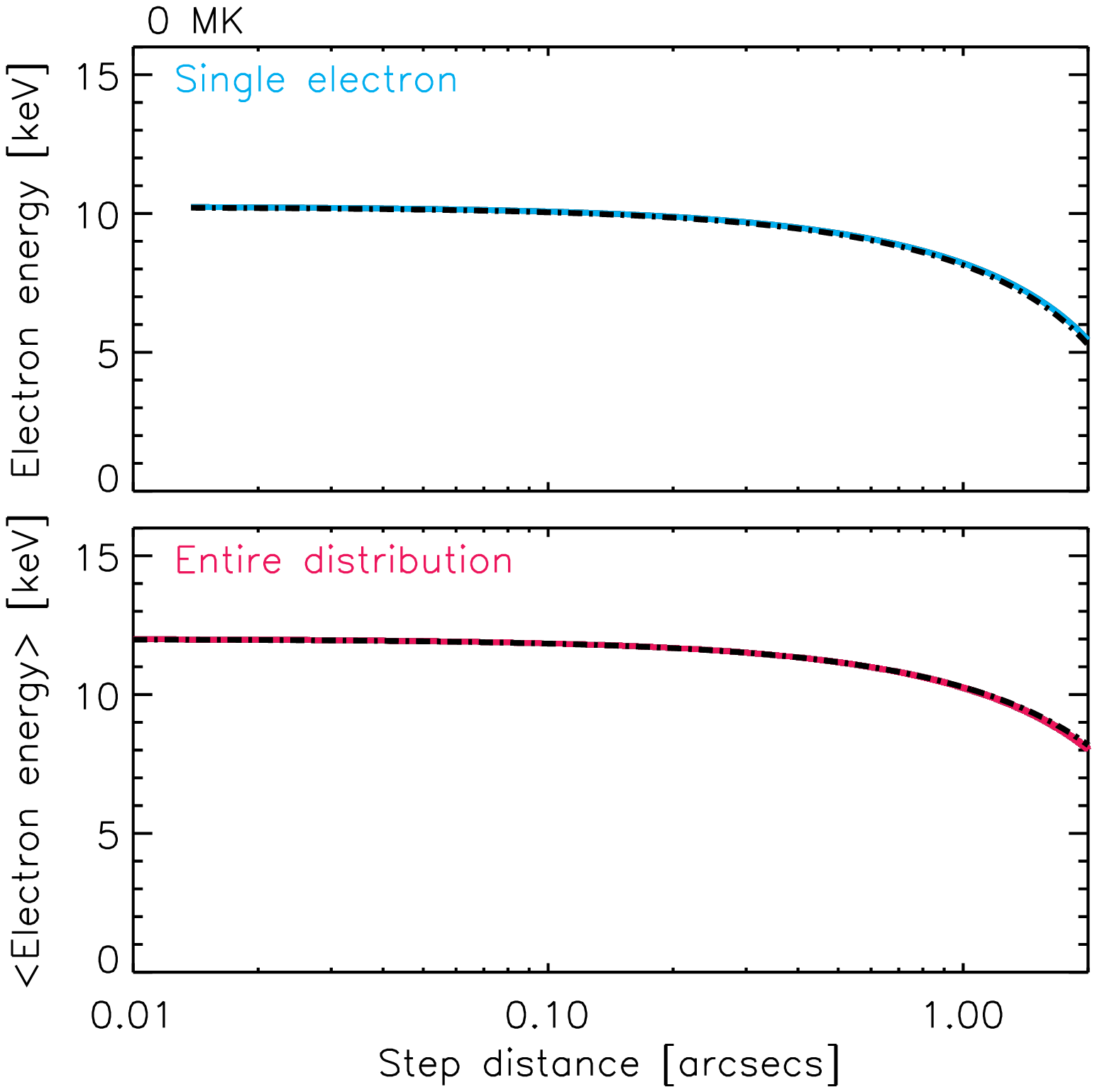}
\includegraphics[width=75mm]{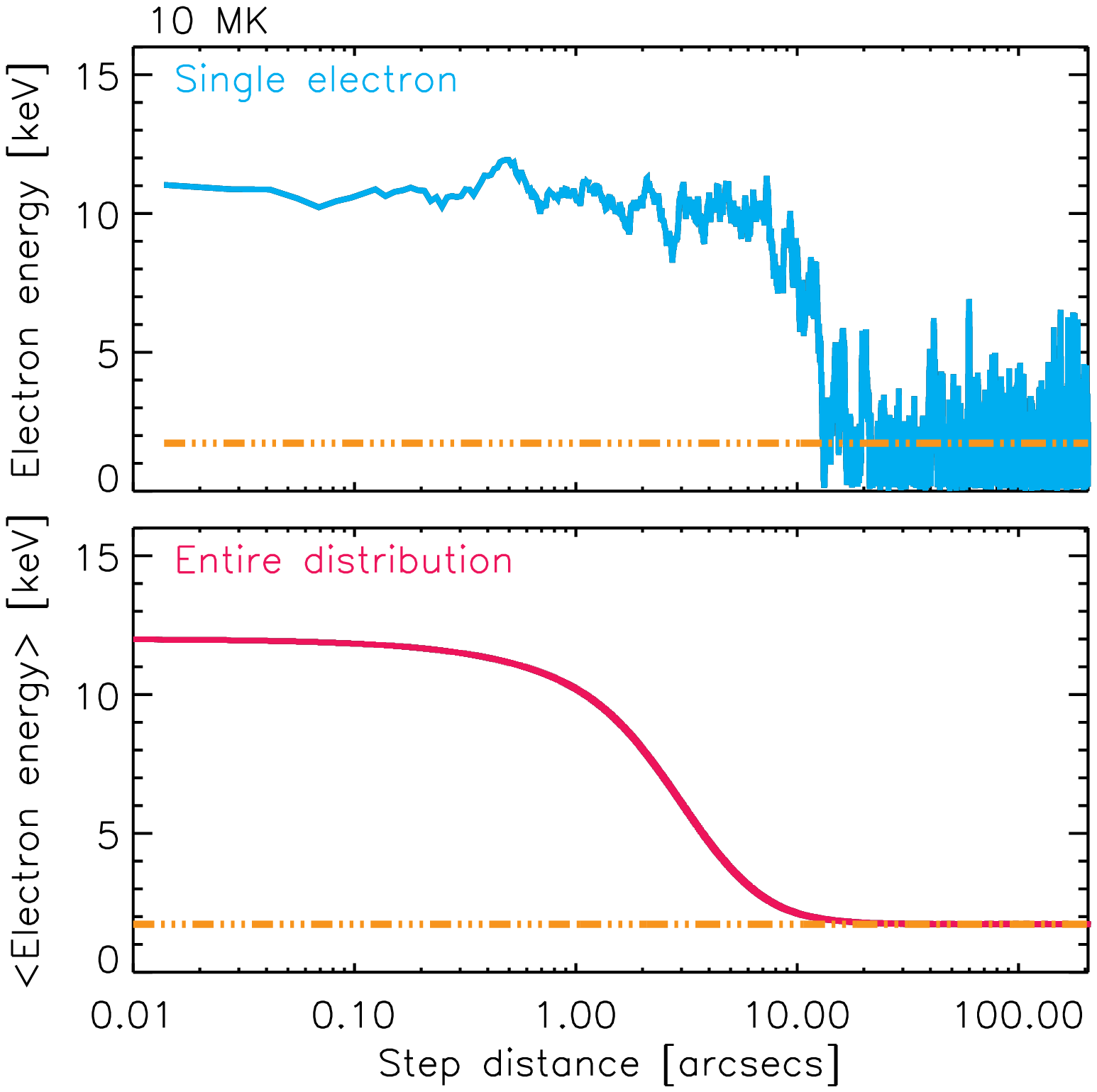}
\includegraphics[width=75mm]{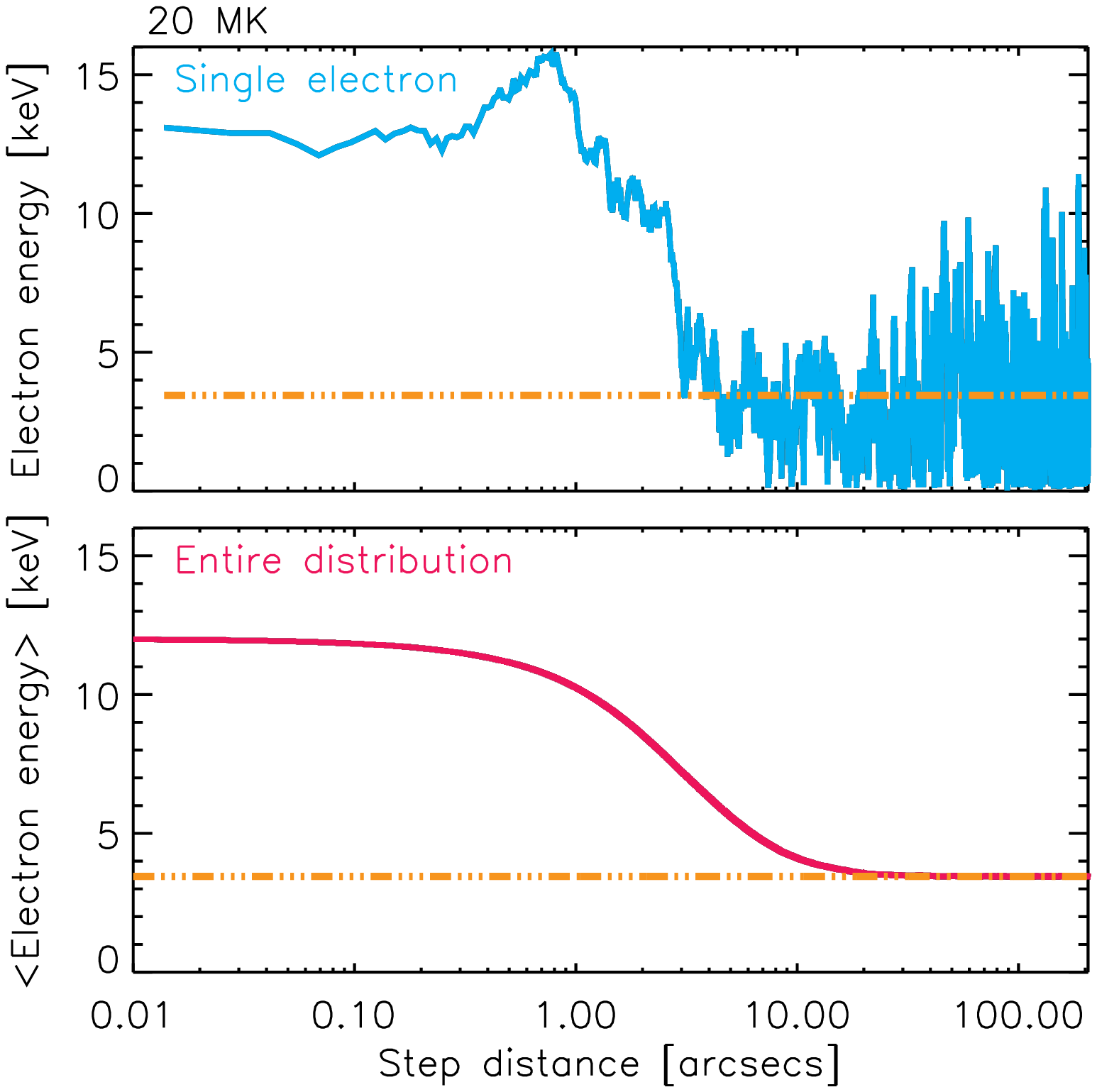}
\includegraphics[width=75mm]{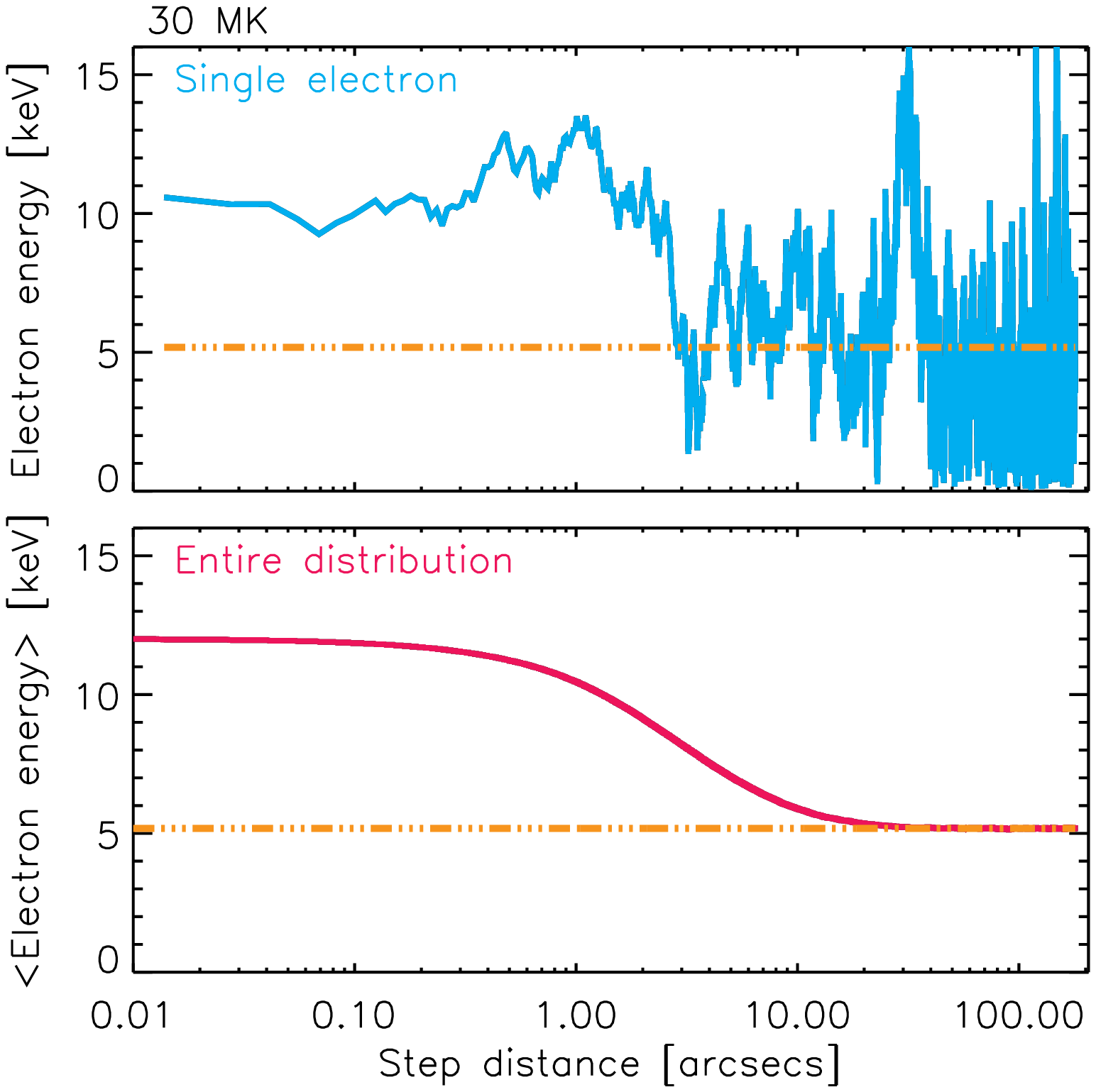}
\caption{{\it Top panels of each plot}: The energy of a single electron for the cold, 10, 20 and 30~MK simulations as a function of the overall path $s$ traveled. For the chosen step-size $\Delta s = 10^5$~cm $\simeq 10^{-3}$ arc seconds, the change in energy over a single step is small. The randomness in the $T$=10, 20 and 30~MK cases is due to thermal fluctuations that increase with plasma temperature; the error associated with this phenomenon is difficult to estimate for a single particle. {\it Bottom panels of each plot}: Average energy of the entire distribution versus the path $s$ traveled, for the  parameters given in Section \ref{sec:input}.  In contrast to the results for a single particle, these show smooth curves, with only small fluctuations for the $T$=10, 20 and 30~MK cases. The black curve indicates the analytical cold-target solution while the orange lines indicate the final average energy of the $F(E)$ distribution.}
\label{fig:check_errors}
\end{figure*}

\subsubsection{Simulation accuracy and limiting cases}

In general, consideration of the errors associated with stochastic simulations are a complex problem and beyond the scope of this paper. However, we can check the convergence of the simulation results against limiting analytical solutions. In the various plots shown in Figure~\ref{fig:check_errors} we plot (top) the energy of a single electron versus the overall step distance traveled and (bottom) the average energy of the entire distribution against the distance traveled.  This was done for $\delta=7$, and for $T$=0, 10~MK, 20~MK and 30~MK. For the cold ($T=0$) case, the error in the energy of a single electron is very small; the stochastic terms in the difference equations~(\ref{eq:sto_x}) through~(\ref{eq:sto_mu}) are negligible and individual electron energies (and hence the average energy of the entire distribution) follow the analytical results very well. However, for a finite temperature target, the stochastic part of the difference equations plays a significant role, the dominance of which increases with $T$. Hence the energy of a single electron fluctuates significantly, especially at low energies.  However, due to ensemble averaging, even for finite target temperatures the average energy of the {\it distribution} exhibits relatively smooth transition from the starting average energy of the distribution to the final average value of the distribution $F(E)$.

\subsection{Numerical results}

\subsubsection{Cold plasma with collisional pitch-angle scattering}

\begin{figure*}
\centering
\includegraphics[width=17cm]{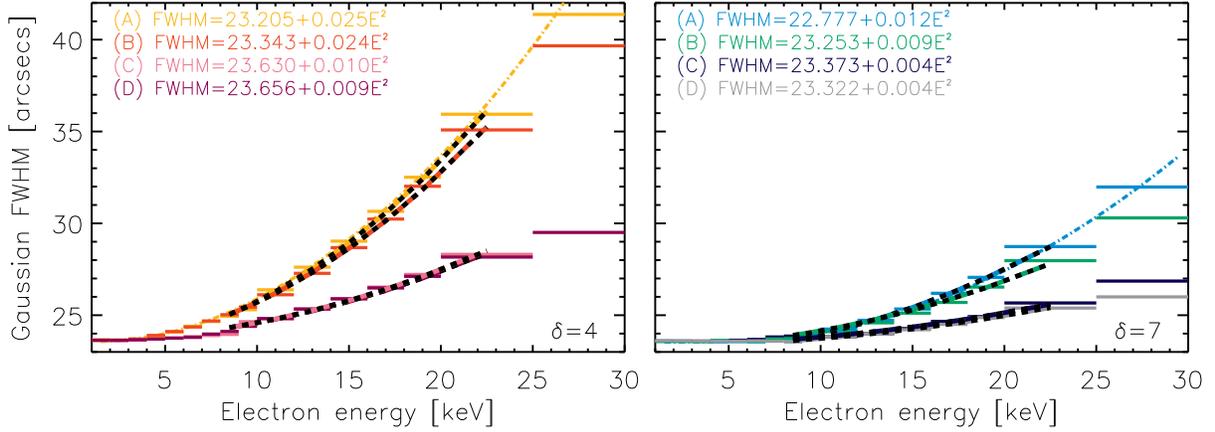}
\caption{Gaussian FWHM versus electron energy $E$ for all cold plasma simulation runs with $n=1\times10^{11}$ cm$^{-3}$ and $\delta=4$ (left plot) and $\delta=7$ (right plot). The cases shown are for: (A) beamed, no pitch-angle scattering (orange, blue), (B) beamed, with pitch-angle scattering (red, green), (C) isotropic, no pitch-angle scattering (pink, navy) and (D) isotropic, with pitch-angle scattering (purple, grey). An equation of the form $L_{0}+\alpha E^{2}$ was fitted to each curve and the values thus found for $L_{0}$ and $\alpha$ are shown on each plot. The fits use energies in the range $\sim 8-25$ keV, matching the energy range fitted to {\em RHESSI} observations. The dashed-dot lines represent the Gaussian FWHM curves fitted to the results of Equation~(\ref{eq:fexgauss}), as in the bottom panels of Figure \ref{fex_good}. As expected, these match well with scenario (A).}
\label{fig:cold_runs}
\end{figure*}

We first consider the case of a cold-target, with different pitch-angle injection and scattering scenarios. Eight simulations were performed, corresponding to two spectral indices ($\delta=4$ and $\delta=7$) and: 

(A) an injected bi-directional beamed distribution of electrons ($\mu=-1,1$) without collisional pitch-angle scattering,

(B) an injected bi-directional beamed distribution of electrons ($\mu=-1,1$) undergoing collisional pitch-angle scattering,

 (C) an initially isotropic pitch-angle distribution of electrons without collisional pitch-angle scattering, and

(D) an initially isotropic pitch-angle distribution of electrons undergoing collisional pitch-angle scattering.

Figure~\ref{fig:cold_runs} shows the Gaussian spatial FWHM plotted against electron energy $E$ for cases (A), (B), (C) and (D), together with fits using Equation (\ref{L}) between $\sim8-25$ keV. This energy range is chosen to match with the energy ranges often used for such observations by {\em RHESSI}. The corresponding values of $\alpha$ and $L_{0}$ for each scenario are shown Figure~\ref{fig:cold_runs}, and there are two general statements that can be made regarding the results. Firstly, the broader the initial pitch-angle distribution, the smaller the source length at a given energy and secondly, the presence of collisional pitch-angle scattering acts to slightly decrease the source length at a given electron energy.  Both effects occur because electrons with $|\mu| < 1$ move a correspondingly smaller distance along the magnetic field. The latter effect is greater at higher electron energies but overall the change is rather small (Figure~\ref{fig:cold_runs}).

The case of an initially isotropic distribution, with or without pitch-angle scattering, produces the flattest (lowest value of $\alpha$) results for each $\delta$. For example, compared with the initially beamed, scatter-free cases for $\delta=4,7$, the isotropic, scatter-free $\alpha$'s are lower by factors of $\sim2.6$ and $\sim3.5$, respectively.

Since the coefficient $\alpha$ (Equation~(\ref{eq: sd2})) in a one-dimensional cold-target formulation is inversely proportional to the ambient density $n$,  the reduced penetration distance associated with the presence of an initially broad pitch-angle distribution and/or collisional scattering will lead to an overestimate of $n$ if the results are interpreted using the one-dimensional cold-target result,  with the exact reduction factor dependent upon the properties of the initial electron distribution and background plasma.

\begin{figure*}
\centering
\includegraphics[width=17cm]{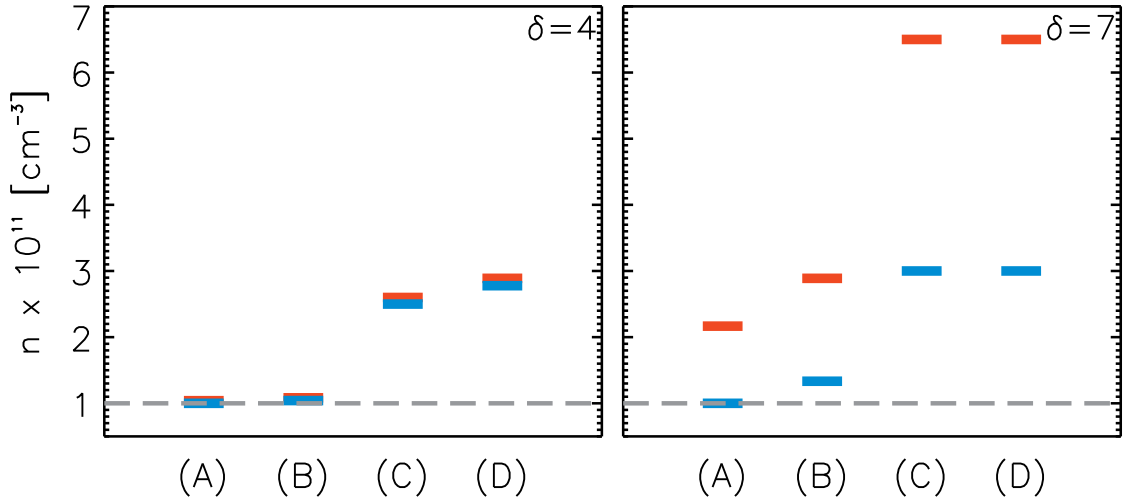}
\caption{For each cold-target simulation scenario -- (A), (B), (C) and (D) -- the value of the coefficient $\alpha$ calculated by fitting each curve in Figure \ref{fig:cold_runs} is used to infer a number density $n$ using two different one-dimensional cold-target approaches:
(1) point injection $\alpha=1/2Kn$ (red) and (2) an extended Gaussian input that is initially beamed with no pitch-angle scattering (blue). The actual number density of $1\times10^{11}$ cm$^{-3}$ is given by the grey dashed line and the inferred value of $n$ is either roughly equal to, or greater than, the actual value.}
\label{fig:find_num_6}
\end{figure*}

We inferred values of $n$ for each of the cases (A), (B), (C) and (D), using two different interpretive approaches:

\begin{enumerate}
\item {$\alpha_{1}=1/2Kn$, i.e., simple one-dimensional propagation within a cold-target, giving $\alpha_{1}=0.026$ arcsecond keV$^{-2}$ for $n=1\times10^{11}$ cm$^{-3}$}
\item {$\alpha_{2}$, found using an extended Gaussian input for an initially beamed distribution with no pitch-angle scattering, i.e., Equation~(\ref{eq:fexgauss}) and scenario (A).  From the lower right panel of Figure~\ref{fex_good}, for $n=1\times10^{11}$ cm$^{-3}$, $\alpha_{2}=0.026$ arcsecond~keV$^{-2}$ for $\delta=4$ and $\alpha_{2}=0.012$ arcsecond~keV$^{-2}$ for $\delta=7$.}
\end{enumerate}

In Figure \ref{fig:find_num_6}, the actual number density of the region $n=1\times10^{11}$ cm$^{-3}$ is shown by the dashed grey line and the values of $n$ inferred from approaches (1) and (2) are shown by the red and blue points, respectively. The inferred number density can be up to six times too large, with the largest effect being for steep spectra (the $\delta=7$ case) and isotropic injection (cases (C) and (D)).

\subsubsection{Hot plasma and collisional pitch-angle scattering}

In this section we study how the effect of a finite-temperature target (in the presence of collisional pitch-angle scattering) changes the electron transport through the plasma and hence the extent of the source with energy. We considered six further simulations corresponding to three target temperatures (10~MK, 20~MK and 30~MK), and pitch-angle scenario (B), an injected beamed electron distribution including pitch-angle scattering, for both $\delta=4$ and $\delta=7$.

\begin{figure*}
\centering
\includegraphics[width=17cm]{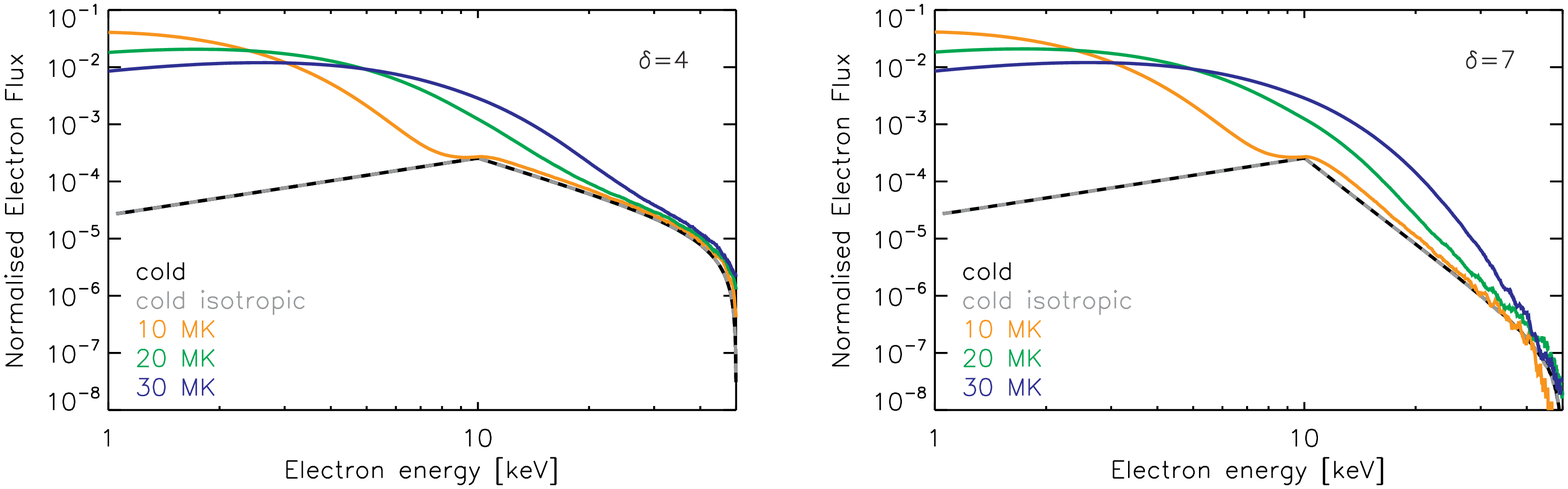}
\includegraphics[width=17cm]{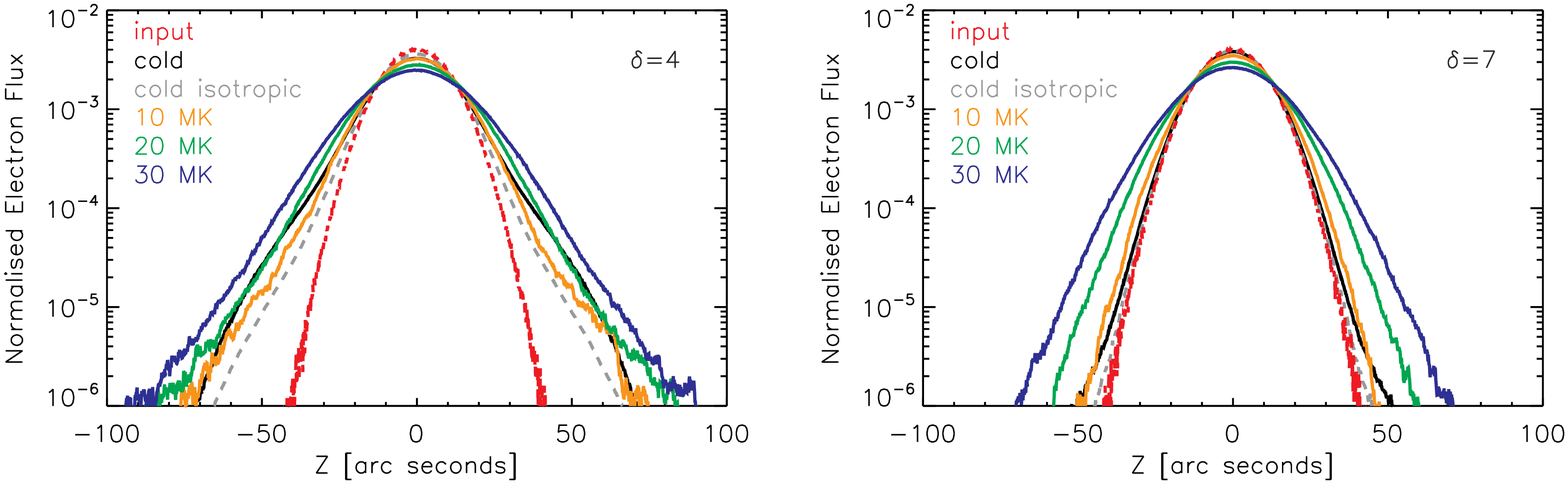}
\caption{{\it Top panels}: spatially-integrated spectra; {\it bottom panels}: energy-integrated spatial distributions for the following scenarios: (1) cold plasma, initially beamed distribution with pitch-angle scattering (black); (2) cold plasma, initially isotropic distribution with pitch-angle scattering (grey); and warm-target cases with (3) $T$=10~MK (orange), (4) $T$=20~MK (green) and (5) $T$=30~MK (blue), with pitch-angle scattering.  Results are shown for both $\delta=4$ (left) and $\delta=7$ (right). }
\label{fig:spectra_x}
\end{figure*}

Figure~\ref{fig:spectra_x} shows both the spatially-integrated spectra and the spectrally-integrated spatial distributions for five different simulations: one-dimensional (beamed) cold-target (black), cold-target with isotropic injection (grey), and beamed injection in three warm-target cases: $T$=10~MK (orange), 20~MK (green) and 30~MK (blue). Figure~\ref{fig:spectra_x} shows only the spatially and spectrally integrated evolutions of the injected electron distribution and does not include the background cold or Maxwellian distribution. The total spatially-integrated spectra are plotted in the top row of panels, for $\delta=4$ (left) and $\delta=7$ (right); the spatial distribution of the spectrally-integrated flux is plotted in the bottom row of panels, again for $\delta=4$ (left) and $\delta=7$ (right).

\begin{figure*}
\centering
\includegraphics[width=17cm]{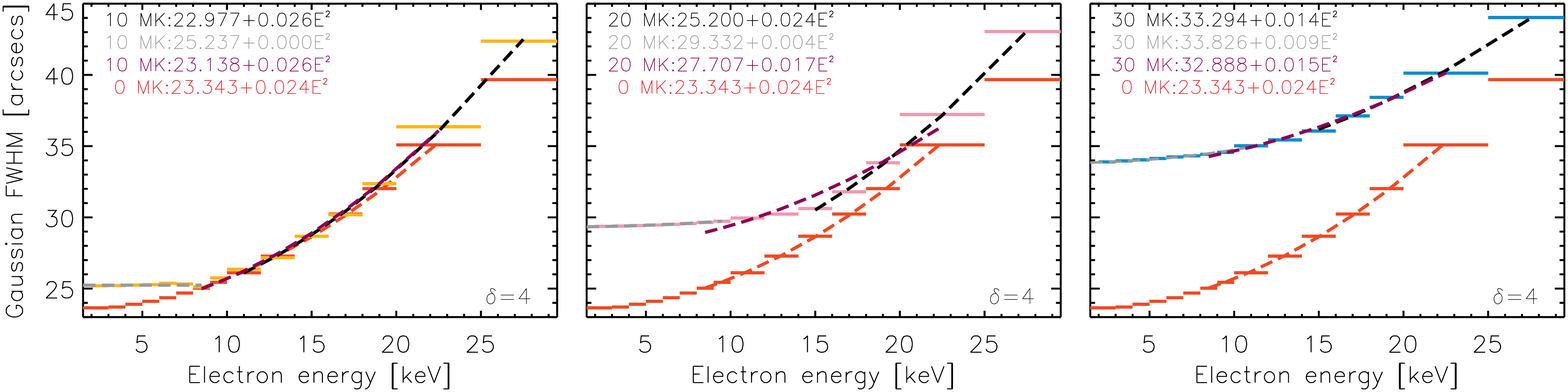}
\includegraphics[width=17cm]{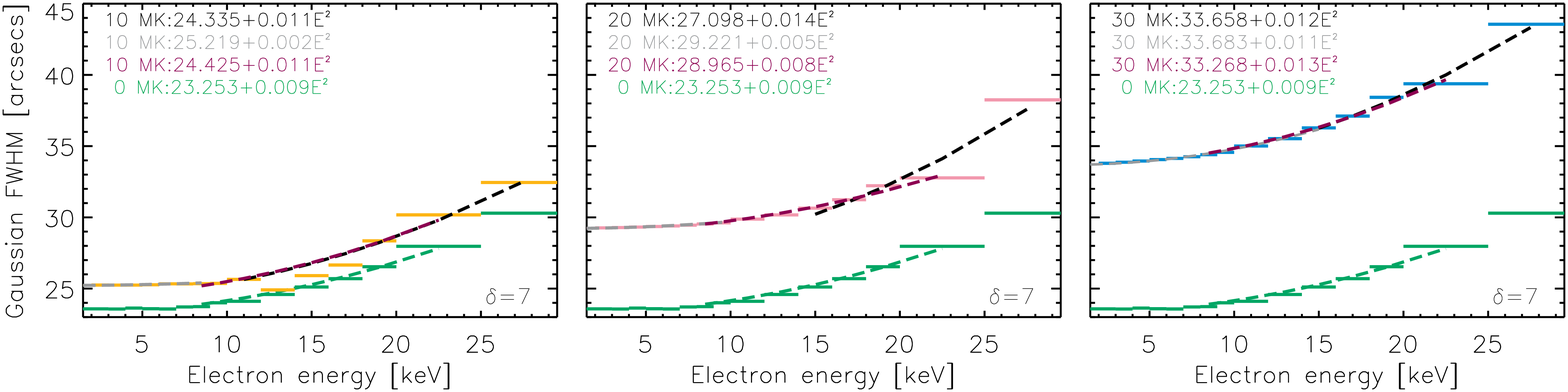}

\caption{Plots of FWHM versus electron energy for $\delta=4$ (top) and $\delta=7$ (bottom), and $T=$ 10 MK, 20 MK and 30 MK (left to right) and $n=1\times10^{11}$ cm$^{-3}$. Fits of the form FWHM=$L_0+\alpha E^{2}$ are shown on each plot. The red and green dashed curves show the corresponding results for the  beamed, cold plasma case with pitch-angle scattering (scenario  (B)).  The purple dashed lines show the best fit in the energy range 8-25~keV, the range used in the fit to {\em RHESSI} observations. Also shown, are the two-component fits, one component representing the thermal diffusion at lower energies (grey dashed curve) and another component representing collisional friction that dominates at higher energies (black dashed curve).}
\label{hot_errors}
\end{figure*}

\begin{figure*}
\centering
\includegraphics[width=17cm]{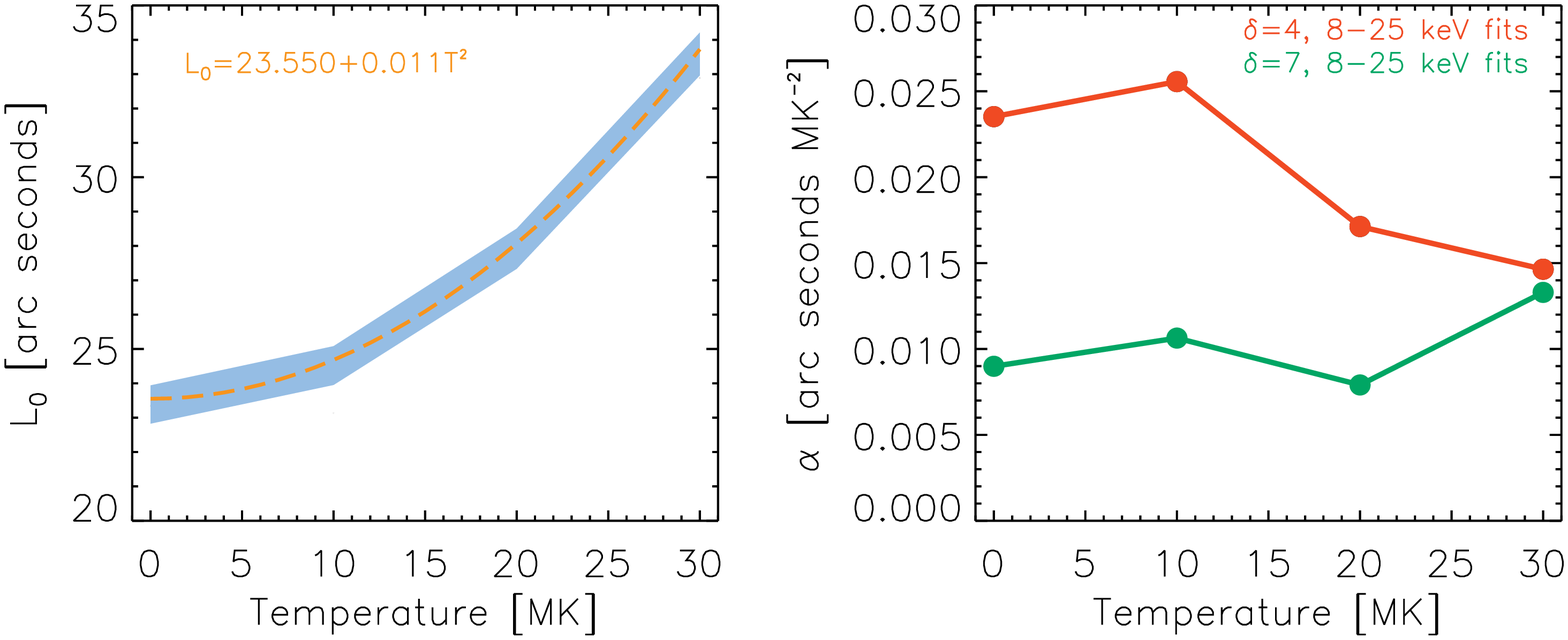}
\caption{{\it Left panel}: $L_0$ versus $T$. The blue band represents the area containing the $L_{0}$ values for both the $\delta=4$ and $\delta=7$ low-energy and 8-25 keV fits from Figure \ref{hot_errors} (grey and purple respectively), plotted against temperature $T$. For each of these curves, a function of the form $L_0=L_{0}(T=0)+\xi T^{2}$ is fitted, and average values of $\langle L_{0}(T=0)\rangle=23\arcsec.5$ and $\langle\xi(n=1 \times 10^{11}$~cm$^{-3})\rangle=0.011$ [arcsec MK$^{-2}$] are found, with $L_0=\langle L_{0}(T=0)\rangle+\langle \xi\rangle T^{2}$ represented by the orange dashed line.  {\it Right panel:} $\alpha$ from the 8-25 keV fits (Figure~\ref{hot_errors}) versus $T$, for $\delta=4$ (red) and $\delta=7$ (green).}
\label{hot_runs_2}
\end{figure*}

Not surprisingly, higher temperature targets tend to make the overall electron spectrum more thermal in form. The lower the temperature of the background Maxwellian plasma, the greater the distinction between the thermal part of the distribution at lower energies and the nonthermal power-law component at higher energies. Also, the inclusion of thermal effects tends to broaden the spatial distribution of the electron distribution, with the effect being more pronounced at higher temperatures.  The spatial spread for a given input distribution is larger for a smaller spectral index because of the larger fraction of higher energy electrons in such flat distributions.  We also found (not shown) that, not surprisingly, the initially beamed distribution (case (B)) shows greater spreads in $z$  than for the same six runs performed for the isotropic injection case (case (D)), see Figure~\ref{fig:spectra_x}.

Figure~\ref{hot_errors} shows the results of the Gaussian fits to the computed spatial distributions for all six warm-target scenarios, together with the corresponding results for the cold-target case.  Compared to the cold-target case, the addition of thermal effects results in changes that affect the inferred values of both $n\propto 1/\alpha$ {\it and} $L_{0}$. Firstly, Figure~\ref{hot_errors} show the inferred acceleration region length increases with temperature;
the magnitude of this increase depends somewhat on the number density $n$ and is relatively independent of the power-law index $\delta$.
This effect is due to the thermal diffusive nature of the electron transport at low energies. This result suggests that the temperature of the background
plasma must be accounted for, when estimating $L_0$ from such observations. The determination of the actual acceleration region length
from the inferred length is discussed further in Section \ref{landn}.

As before, we fit curves of the form (\ref{L}) between $\sim8-25$~keV, for a better comparison with observations.  These are shown by the purple
dashed lines and the values of $L_0$ and $\alpha$ from the purple fits are shown on each panel of Figure~\ref{hot_errors}.
However, the presence of a finite background temperature causes the lower energies of the distribution, in particular, to be dominated
by thermal diffusion and hence analysis of the curves in Figure~\ref{hot_errors} shows that overall, the FWHM over the entire plotted energy
range is not so well-fitted by a single curve of the form FWHM$(E)=L_0+\alpha E^{2}$. This can be clearly seen for the 20 MK, $\delta=4$ curve.
Therefore, we chose to fit the results with two other FWHM$(E)=L_0+\alpha E^{2}$ curves; one component representing the lower energy
values that are controlled mainly by thermal diffusion (grey curve) and another component representing higher energies mainly controlled
by collisional friction, since the FWHM values should return to match those of a cold-target case when $E\gg k_{B}T$.
The $L_0$ and $\alpha$ values found from the grey and black curves are also shown
on each panel of Figure~\ref{hot_errors}.

To illustrate, for the $T$=10~MK case, the FWHM values match those of the cold case (red or green dashed line) after $\sim10$~keV, for both the $\delta=4$ and $\delta=7$ cases. This is because the temperature diffusion is limited to energies below $\sim8$~keV (grey curve); Figure~\ref{hot_errors} clearly shows this transition. Therefore for the 10 MK case, our $8-25$~keV fits (purple) match that of the higher energy black fits and cold cases reasonably well for both $\delta=4$ and $\delta=7$. By $T$=20~MK, the energy range between $8-25$ keV is not so well fitted by a curve of the form of Equation (\ref{L}) and occurs because the trend of the FWHM moves from being dominated by the effects of thermal diffusion to being dominated by the effects of collisional friction at approximately $15$ keV, right in the middle of the range we are using for the fit. This is clear for the $\delta=4$ case but harder to see for $\delta=7$ case due to the smaller values of $\alpha$. The $\alpha$ values of the friction-dominated fits (black curves) are only approximately the same as for the cold plasma case after $\sim17$ keV. Also the diffusion at 20~MK noticeably influences the length values at all energies plotted, with the FWHM values above $\sim$17~keV lying above those for the cold case. By $T$=30~MK, the entire plotted energy range and our fitted energy range between $8-25$ keV is mainly controlled by thermal diffusion and the $\alpha$ values for both the $\delta=4$ and $\delta =7$ cases are similar. All plotted FWHM values are much larger than that of equivalent cold cases, over $10\arcsec$ at $1$ keV. For the $8-25$ keV fits, the $\delta=4$ value is smaller than that of the equivalent cold case, and the $\delta=7$ value is slightly larger.

\subsubsection{Inferring the acceleration region length $L_0$ and density $n$}\label{landn}

The thermal diffusion-component (grey dashed) curves in Figure~\ref{hot_errors} use Equation~(\ref{L}) to fit the FWHM values at lower energies, and hence give us $L_0$, the inferred length of the acceleration region. For a given temperature, the values of $L_0$ found for both $\delta=4$ and $\delta=7$ are approximately the same, with an average value of $25\arcsec$ for $T$=10~MK, $29\arcsec$ for $T$=20~MK and $34\arcsec$ for $T$=30~MK.

Figure \ref{hot_runs_2} (left) plots the values of $L_0$ found for the thermal diffusion-dominated (grey curve) and $8-25$ keV fits against $T$. Each is fitted with a curve of the form

\begin{equation}
L_0(T,n)=L_{0}(T=0)+\xi(n) T^{2}=23\arcsec.5+\xi(n) T^{2} \,\,\, .
\label{eq:LT}
\end{equation}
By fitting Equation (\ref{eq:LT}) to each, $\xi$ is found for both ``global'' and thermal diffusion-dominated fits, and an average value of $\langle\xi(n=1\times10^{11})\rangle=0.011$ arcsecond MK$^{-2}$ is calculated averaging the four fits.

To summarize, if the size $L_0(T=0)$ and number density $n$ of the region have been inferred from a cold-target analysis, and $n$ is close to $n=1\times10^{11}$ cm$^{-3}$, then  the actual extent of the acceleration region is less than would be inferred using a cold-target formula. Quantitatively, the actual size of the acceleration region $L_0$ can be approximated by the expression

\begin{equation}\label{T-correction}
L_0=L_0(T=0)-0.011 \, T^{2} \,\,\, ,
\end{equation}
where $L_0(T=0)$ is the value deduced from a fit using the cold-target formula to an observation.

The right panels in Figure~\ref{hot_runs_2} also show how $\alpha$ from the $8-25$ keV fits changes with $T$ for both $\delta=4$ and $\delta=7$. For $\delta=4$, $\alpha$ decreases between $T$=10~MK and $T$=30~MK.  This is expected, since for higher temperatures, particle diffusion is controlling the shape of the curve and hence the $\delta=4$ cold-target case has a relatively high $\alpha$ value. However this is not the case for $\delta=7$, where between 10-30 MK we see $\alpha$ increasing with $T$.

From the plots in Figure \ref{hot_runs_2}, the values of $\alpha$ for the fits between 8-25 keV can be used to infer a number
density from observations. Two cold-target approaches are used: (1) $\alpha={1}/2Kn$, and (2) an extended source Gaussian input as found from Equation (\ref{eq:fexgauss}). Also, using our results from the cold-plasma cases we can expand (2) to account for the initial beaming of the distribution so that a range of $n$ can be found.  Finally, we can use (3), which is the same as (2) but accounts for pitch-angle scattering.

\begin{figure*}
\centering
\includegraphics[width=17cm]{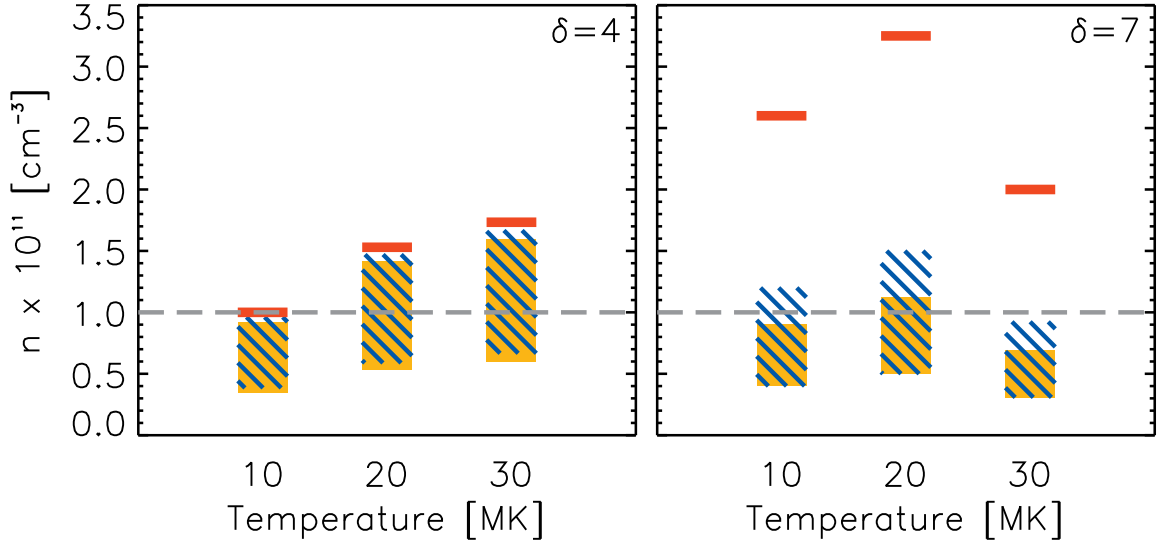}
\caption{Cold plasma fits are applied to the different hot plasma simulation curves to determine an inferred density that can be compared with the actual density of the region: (1) (red lines): $\alpha_{1}={1}/2Kn$. (2) (blue hashed areas): an extended Gaussian injection model with no pitch-angle scattering that is initially either beamed (Equation~(\ref{eq:fexgauss})) or isotropic (found from the cold plasma simulation -- see Figure~\ref{fig:cold_runs}). (3) (orange regions): as for (2), but with collisional pitch-angle scattering included.  For both (2) and (3), the highest inferred values for $n$ are for a completely beamed distribution.}
\label{new_simn}
\end{figure*}

The inferred values of $n$ for $T=10, 20, 30$~MK and for $\delta=4,7$ are shown in Figure~\ref{new_simn}. For $\delta=4$ the largest inferred value  is $\sim1.7$ times larger than the actual density and the smallest is around three times smaller; for $\delta=7$, the largest value is $\sim3.3$ times larger and the smallest value is again about 3 times smaller. In general, (1) (cold-target, point injection, red lines) produces the largest differences, which is not surprising since our input was an extended Gaussian, rather than point-injection, source. However, even this simple analytical case, that accounts very poorly for the true physical properties of the electron distribution, only increases the number density by a factor of about $3$ (for a beamed finite temperature case). (2) and (3) (extended injection models, without and with collisional scattering in the target, respectively) that do not account for the finite temperature of the plasma provide an inferred value for $n$ that is quite close to the true value of $n$, with the biggest uncertainty due to the unknown degree of beaming of the injected distribution.

\section{Discussion and conclusions}\label{discussion}

Our aim in this paper was to understand how the different injected pitch-angle distributions, the collisional pitch-angle scattering, and a finite target temperature change electron transport through a plasma and hence the spatial properties of compact hard X-ray sources in solar flares.

Our simulations show three main results:

\begin{enumerate}

\item{Collisional pitch-angle scattering alone does not dramatically change the behavior
of source length with electron energy.}

\item{Beaming of the initial electron pitch-angle distribution does produce a significant change in the variation
of the length of the X-ray source with energy; distributions that are initially beamed produce a larger variation of length
with energy, a consequence of the fact that the collisional stopping distance is now projected onto the direction defined
by the guiding magnetic field.  The difference in the coefficient $\alpha$ can be up to a factor of $6$ if a beamed approximation is used for a distribution that is in fact completely isotropic. The uncertainty in the initial angular distribution of the injected electrons produces the largest uncertainty in the inferred number density $n$.}

\item{The finite temperature of the target atmosphere leads to thermal diffusion and an increase of the inferred acceleration region length. The FWHM versus energy consists of two competing components, one due to diffusion that is dominant at lower energies, and another due to collisional friction that is dominant at higher energies.  Which component predominates depends in a complicated way on the temperature of the region, on the density $n$, and even on the spectral index $\delta$. Therefore the use of a cold-target approximation with a single fitted curve to infer properties of the acceleration region should always be used with caution. Our results show that applying a cold model to a warm plasma changes the inferred acceleration length $L_0$ by several arc seconds (see Equation~(\ref{T-correction})) and the inferred number density by up to a factor of $3$ (in either direction), depending mainly on the initial beaming of the electron distribution (see Figure~\ref{new_simn}).}
\end{enumerate}

The influence of the effects studied in this paper also influence the determination of other quantities, such as the acceleration region filling factor $f$ (the fraction of the apparent source volume in which acceleration occurs) and the specific acceleration rate (the fraction of the ambient electron population that is accelerated per unit time).  The filling factor $f$ is defined by

\begin{equation}\label{filling-factor}
f = \frac{EM}{n^2 V} \,\,\, ,
\end{equation}
where $V = (\pi W^2/4) L_0$ is the volume of the acceleration region, determined from the inferred value of $L_0$ and the observed lateral extent $W$ of the (cylindrical) acceleration volume, and the emission measure $EM$ is determined from, for example, fits to the spatially-integrated soft X-ray spectrum of the flare.  The effects studied in this paper show that in general application of a one-dimensional cold-target formula leads to erroneously high inferred values for both the acceleration region length $L_0$ (see Figure~\ref{hot_errors}) and density $n$ (see Figure~\ref{new_simn}).  Use of such erroneously high values of $L_0$ and $n$  leads to an overestimate of the denominator in Equation~(\ref{filling-factor}) and so an underestimate of the filling factor $f$.

In a study of 24 coronal thick-target events, using the one-dimensional cold-target result~(\ref{cold_theory}) to estimate $L_0$ and $n$, \citet{2013ApJ...766...28G} found filling factors $f$ that were generally somewhat less than unity. The results of this paper therefore lend support to a value of $f$ being even closer to unity than previously thought. Indeed, given that $f$ cannot exceed unity, this may place constraints on the allowable values of $n$ and $L_0$.  And, since the inferred values of $n$ depend significantly on the pitch-angle distribution of the injected electrons, this could conceivably be used to constrain the form of the injected pitch-angle distribution.  In particular, broad injected distributions lead to relatively small values of the coefficient $\alpha$ (see Figure~\ref{fig:cold_runs}) and hence to inferred densities that are higher than the actual target density (Figure~\ref{new_simn}). Correcting for such an effect in the interpretation of a particular event could imply an actual target density that was too small to be compatible with the observationally-inferred emission measure, thus ruling out the hypothesis of a broad injected distribution of accelerated electrons.

Inference of the acceleration region length $L_0$, lateral extent $W$, and density $n$ also gives the number of electrons available for acceleration:

\begin{equation}\label{number_of_particles}
{\cal N} = n \, V = n \, \left ( \frac{\pi W^2}{4} \right ) \, L_0 \,\,\, .
\end{equation}
This, combined with the inference of $d{\cal N}(E_0)/dt$, the rate of electron acceleration beyond energy $E_0$ (obtained rather straightforwardly from spatially-integrated hard X-ray data) provides the value of the {\it specific acceleration rate} (electrons~s$^{-1}$ per ambient electron)

\begin{equation}\label{acceleration-rate}
\eta(E_0) = \frac{1}{\cal{N}} \, \frac{d {\cal N}(E_0)}{dt} \,\,\, .
\end{equation}
Overestimating the value of the acceleration region volume and density through the use of an over-simplistic one-dimensional cold-target model thus causes an overestimate of ${\cal N}$ and, since $d{\cal N}(E_0)/dt$ is fixed, this causes an underestimate of $\eta(E_0)$.  In their multi-event study, \citet{2013ApJ...766...28G} found typical values for $\eta(E_0=20~{\rm keV})$ were of order $10^{-2}$~s$^{-1}$; application of the more physically realistic source models considered herein will increase $\eta$ even further, thus placing more profound constraints on the electron acceleration mechanism.

 For future work, it should be noted that our simulations could be made more self-consistent by allowing for the temperature increase of the background plasma due to the energy loss of the injected electron distribution. They also could be augmented by including spatial variations in temperature and/or density along the loop.  Also, it should be noted that a recent study by \citet{2014ApJ...780..176K} shows how the presence of {\it non-collisional} pitch-angle scattering (e.g., involving magnetic field inhomogeneities) results in a different (non-quadratic) predicted behavior for the variation of source length with energy. The code we have developed for this work can be rather straightforwardly extended to the study of diffusion of particles across the guiding field in a warm-target \citep[e.g.,][]{2011A&A...535A..18B} and hence to study the variation of source length with energy in this alternative scenario.

Finally, this study has shown that simulating more realistic effects such as isotropic electron distributions \citep[as indicated by recent studies  -- e.g.,][]{2006ApJ...653L.149K,2007A&A...466..705K,2013SoPh..284..405D} in general produces a more gradual variation of source length with energy, i.e.,  smaller values of the coefficient $\alpha$. Therefore, depending on the electron distribution spectral index, observed steep behaviors (high values of $\alpha$) may be indicative of other processes at work within the coronal region. For instance, throughout our simulations we assumed that the length of the acceleration region length $L_0$ did not depend on electron energy $E$. However, depending on the acceleration process, this may not be the case. For example, if $L_0$ grows with energy, this may produce a larger value of $\alpha$ than expected and hence the analysis of this effect may tell us something about the acceleration mechanism itself.

\acknowledgments

We thank the referee for the useful comments that helped to improve the text of this paper. NLSJ is funded by STFC and a SUPA scholarship. EPK and NHB gratefully acknowledge the financial support by STFC Grant, and by the European Commission through the HESPE (FP7-SPACE-2010-263086) Network. AGE was supported by NASA Grant NNX10AT78J.

\bibliographystyle{apj}
\bibliography{refs_rhessi}

\appendix

\section*{Appendix:  Fokker-Planck Equation Coefficients}\label{appA}

In order to describe the transport of electrons through a coronal plasma of finite temperature $T$, accounting for collisional pitch-angle scattering, a Fokker-Planck equation can be used. We take the three-dimensional form from \citet[e.g.,][]{1981phki.book.....L,1986CoPhR...4..183K} in spherical coordinates. Assuming azimuthal symmetry and adding a source term for electrons $S(v,\mu,z)$, this is given by

\begin{equation}\label{eq:fp}
\frac{df(v,z,\theta,t)}{dt}=\frac{\partial f}{\partial t}+v\cos\theta \frac{\partial f}{\partial z}=-\frac{1}{v^{2}} \, \frac{\partial}{\partial v}
\left(v^{2} \, S_{v}\right) - \frac{1}{v\sin\theta}\frac{\partial}{\partial \theta}
\left(\sin\theta \, S_{\theta}\right) +S(v,\mu,z),
\end{equation}
where $f(v,\theta,z,t)$ is the phase-space distribution function (electrons~cm$^{-3}$~{[cm~s$^{-1}$]}$^{-3}$), $v$ (cm~s$^{-1}$) is the total particle speed, $\theta$ is the pitch-angle to the guiding magnetic-field (along the direction $z$), $t$ is time (s) and the coefficients $S_{v}$ and $S_{\theta}$ are given by

\begin{equation}\label{eq:SS}
S_{v}=-D_{vv} \, \frac{\partial f}{\partial v} + F_{v} \, f
 \,\,\, , \qquad S_{\theta}=-D_{\theta\theta} \, \frac{1}{v} \, \frac{\partial f}{\partial \theta} \,\,\, .
\end{equation}
Here $D_{vv}$ and $D_{\theta\theta}$ are the velocity and pitch-angle diffusion terms while $F_{v}$ is the velocity collisional friction term. These three terms are respectively given by

\begin{eqnarray}\label{eq:dcvv}
D_{vv}=\frac{\Gamma}{2v} \, \left(\frac{{\rm erf}(u)}{u^{2}
}-\frac{{\rm erf}^{'}(u)}{u}\right) &\equiv & \frac{\Gamma}{v} \, G(u)
\cr
D_{\theta\theta}=\frac{\Gamma}{4v}\left( \left[2-\frac{1}{u^{2}}\right]{\rm erf}(u)
+ \frac{{\rm erf}^{'}(u)}{u} \right) &\equiv & \frac{\Gamma}{2v}\biggl ( {\rm erf}(u)-G(u) \biggr )
\cr
F_{v}=-\frac{\Gamma}{v^{2}}\left({\rm erf}(u)-u \, {\rm erf}^{'}(u)\right)
&\equiv & -\frac{2 \, \Gamma}{v^{2}} \, u^{2} \, G(u) \,\,\, ,
\end{eqnarray}
where the dimensionless velocity $u=v/(\sqrt{2} \, v_{th})$, $v_{th}=\sqrt{k_{B}T/m_{e}}$ , $\Gamma=4\pi e^{4} \ln\Lambda \, n /m_{e}^{2}$, ${\rm erf}(u)$ is the error function and $G(u)$ is the Chandrasekhar function, given by

\begin{equation}\label{eq:gcha}
G(u)=\frac{{\rm erf}(u)-u \, {\rm erf}^{'}(u)}{2u^{2}} \,\,\, .
\end{equation}
Substituting into the Fokker-Planck equation~(\ref{eq:fp}) gives

\begin{eqnarray}\label{eq:fpv}
\frac{d f(v,\theta,t)}{d t}
&= & \frac{\Gamma}{2v^{2}} \left \{ \frac{\partial}{\partial v}\left(2 \, v \, G(u) \, \frac{\partial f(v,\theta,t)}{\partial v}
+4 \, G(u) \, u^{2} \, f(v,\theta,t)\right) + \right . \cr
&+ & \left . \frac{1}{v\sin\theta} \, \frac{\partial}{\partial \theta}\left(\sin\theta \, \biggl [ {\rm erf}(u) -G(u) \biggr ] \, \frac{\partial f(v,\theta,t)}{\partial \theta} \right) \right \} \,\,\, .
\end{eqnarray}

Current imaging spectroscopy X-ray observations with instruments such as {\em RHESSI} have a time resolution the order of several seconds (it takes a full spacecraft rotation period $\sim$4 s to yield a reliable image), which is much longer than the timescale for transport of deka-keV electrons ($v \sim 10^{10}$~cm~s$^{-1}$) along the typical length of a coronal loop ($\sim 10^9$~cm).  Therefore, it is appropriate to consider the time-independent case. It is also convenient to convert from the variable $\theta$ to the variable $\mu=\cos{\theta}$, giving

\begin{eqnarray}\label{eq:fpvxmu}
\mu \, v \, \frac{\partial f(v,\mu,z)}{\partial z}
&= & \frac{\Gamma}{2v^{2}} \left \{ \frac{\partial}{\partial v}\left(2 \, v \, G(u) \, \frac{\partial f(v,\mu,z)}{\partial v}
+4 \, u^{2} \, G(u) \, f(v,\mu,z)\right) + \right . \cr
&+ & \left . \frac{1}{v} \, \frac{\partial}{\partial \mu}\left( (1-\mu^2) \, \biggl [ {\rm erf}(u) -G(u) \biggr ] \, \frac{\partial f(v,\mu,z)}{\partial \mu} \right) \right \} + S(v,\mu,z) \,\,\, .
\end{eqnarray}

We assume that the source term $S(v,\mu,z)$ is separable in $v,\;\mu$ and $z$, with the spatial variation assumed to have a Gaussian form:

\begin{equation}\label{eq:source_v}
S(v,\mu,z)=f_{0}(v) \, \frac{1}{\sqrt{2\pi d^{2}}}\exp{\left(-\frac{z^{2}}{2d^{2}}\right)} \, H(\mu) \,\,\, ,
\end{equation}
where $f_{0}(v)$ and $H(\mu)$ are the initial velocity and pitch-angle distribution functions.

For a background plasma with a finite temperature $T$, the input distribution will evolve to a thermal distribution of the form

\begin{equation}\label{eq:fv_the}
f(v)\sim\exp{\left(-\frac{mv^{2}}{2k_{B}T}\right)} \,\,\, ,
\end{equation}
leading to an average kinetic energy of

\begin{equation}\label{eq:fv_kT}
\left\langle\frac{mv^{2}}{2}\right\rangle
=\frac{\int_{0}^{\infty}\frac{mv^{2}}{2}f(v)d^{3}v}{\int_{0}^{\infty}f(v)d^{3}v}
=\frac{3}{2} \, k_{B}T \,\,\, .
\end{equation}

In the high electron velocity limit $u\gg1$, { \bf ${\rm erf} (u)\rightarrow 1$ and one finds $G(u)\rightarrow 1/2u^2=(v_{th}/v)^{2}$}. In this limit Equation~(\ref{eq:fpvxmu}) becomes

\begin{eqnarray}\label{eq:fpvxmu_he}
\mu \, v \, \frac{\partial f(v,\mu,z)}{\partial z}
&= & \frac{\Gamma}{v^2} \left \{ \frac{\partial}{\partial v} \left( \frac{v_{th}^2}{v} \frac{\partial f(v,\mu,z)}{\partial v} \,
+ \, f(v,\mu,z)\right) + \right . \cr
&+ & \left . \frac{1}{2v}
\frac{\partial}{\partial \mu} \left( (1-\mu^2) \, \frac{\partial f(v,\mu,z)}{\partial \mu} \right) \right \}  + S(v,\mu,z)\,\,\, .
\end{eqnarray}
If the temperature of the plasma is also small compared to the typical particle energies, then we can formally take $T=0$ (i.e., $v_{th}=0$).  Equation~(\ref{eq:fpvxmu_he}) then becomes

\begin{equation}\label{eq:fpvxmu_noT}
\mu \, \frac{\partial f(v,\mu,z)}{\partial z}
= \frac{\Gamma}{v^3} \, \left \{ \frac{\partial f(v,\mu,z)}{\partial v} + \frac{1}{2 v} \, \frac{\partial}{\partial \mu} \left( (1-\mu^2) \, \frac{\partial f(v,\mu,z)}{\partial \mu} \right) \right \}  + S(v,\mu,z)\,\,\, ,
\end{equation}
which is the transport equation for a cold plasma with azimuthal symmetry, often used in solar physics \citep[e.g.,][]{1981SvA....25..215K}.

The energy flux $F(E,\mu,z)$ (electrons~cm$^{-2}$~s$^{-1}$~erg$^{-1}$) is related to the three-dimensional phase-space distribution function $f(v,\mu,z)$ by

\begin{equation}\label{con_fF1}
v \, f(v,\mu,z) \, v^2 \, dv = F(E,\mu,z) \, dE \,\,\, ,
\end{equation}
so that

\begin{equation}\label{con_fF}
f(v,\mu,z)= \frac{m}{v^{2}} \, F(E,\mu,z) = \frac{m^{2}}{2} \, \frac{F(E,\mu,z)}{E} \,\,\, .
\end{equation}
Using this relation, we can write the Fokker-Planck equation~(\ref{eq:fpvxmu}) in terms of electron energy $E$ and the electron flux distribution $F(E,\mu,z)$, which is a more useful form for comparison with observations.  The result is

\begin{eqnarray}\label{eq: fp_e}
\mu \, \frac{\partial F}{\partial z} &= & \Gamma m^2 \left \{ \frac{\partial}{\partial E} \left[ G (u[E] ) \, \frac{\partial F}{\partial E} + \frac{G (u[E] )}{E} \, \left ( \frac{E}{k_B T}-1 \right ) \, F \right] + \right . \cr
&+ & \left . \frac{1}{8E^2} \, \frac{\partial}{\partial \mu} \left [ (1-\mu^{2}) \biggl ( {\rm erf} (u[E] ) - G (u[E] ) \biggr ) \, \frac{\partial F}{\partial \mu} \right ] \right \} + S_F(E,\mu,z)\,\,\,\, ,
\end{eqnarray}
where we have used $u(E)=\sqrt{E/k_B T}$.

The solar corona contains elements other than hydrogen, and for an element with atomic number $\zeta$, the Coulomb energy loss scales as $\zeta^2$ \citep[e.g.,][]{1978ApJ...224..241E}. We thus account for these additional elements by adopting an effective atomic number $\zeta_{eff}=\sum_{i} n_{i} \zeta_{i}^{2}/\sum_i n_i$. Defining $\Gamma_{eff}=\Gamma \zeta_{eff} m^{2}$, we obtain

\begin{eqnarray}\label{eq: fp_e_zeff}
\mu \, \frac{\partial F}{\partial z} &= & \Gamma_{eff} \left \{ \frac{\partial}{\partial E} \left[ G (u[E] ) \, \frac{\partial F}{\partial E} + \frac{G (u[E] )}{E} \, \left ( \frac{E}{k_B T}-1 \right ) \, F \right] + \right . \cr
&+ & \left . \frac{1}{8E^2} \, \frac{\partial}{\partial \mu} \left [ (1-\mu^{2}) \biggl ( {\rm erf} (u[E] ) - G(u[E] ) \biggr ) \, \frac{\partial F}{\partial \mu} \right ] \right \}  + S_F(E,\mu,z)\,\,\,\, .
\end{eqnarray}

\end{document}